\documentclass[a4paper,11pt]{article}
\pdfoutput=1 
\usepackage{jcappub} 
\usepackage[T1]{fontenc} 
\usepackage[utf8x]{inputenc}
\usepackage{multirow}
\usepackage{booktabs}
\usepackage{bm}
\usepackage{times}
\usepackage{tensor}
\usepackage{graphicx}  
\usepackage{color}
\usepackage{times}
\usepackage{inputenc}
\usepackage{float}
\usepackage{bm}
\usepackage{ulem}
\usepackage{multirow}
\usepackage{url}
\usepackage{natbib}
\usepackage{mathrsfs}
\usepackage{physics}
\usepackage{comment}
\usepackage[table,xcdraw]{xcolor}
\usepackage{booktabs,makecell}
\usepackage{comment}
\usepackage{amsmath}
\usepackage{pifont}
\usepackage{rotating}
\usepackage[colorlinks=true,citecolor=blue,urlcolor=blue,linkcolor=blue]{hyperref}
\title{Dark Matter Influence on Quarkyonic Stars: A Relativistic Mean Field Analysis}
\author[a,b]{D. Dey,}
\author[c,1]{Jeet Amrit Pattnaik \note{Corresponding author,}} 
\author[d]{H. C. Das,}
\author[e]{Ankit Kumar,}
\author[c]{R. N. Panda,} 
\author[a,b]{S. K. Patra}

\affiliation[a]{Institute of Physics, Sachivalaya Marg, Bhubaneswar-751005, India.}
\affiliation[b]{Homi Bhabha National Institute, Training School Complex, 
Anushakti Nagar, Mumbai 400094, India.}
\affiliation[c]{Department of Physics, Siksha $'O'$ Anusandhan, Deemed to be University, Bhubaneswar -751030, India.}
\affiliation[d]{Istituto Nazionale di Fisica Nucleare, Sezione di Catania,
Dipartimento di Fisica, Universit\'a di Catania, Via Santa Sofia 64, 95123 Catania, Italy.}
\affiliation[e]{Key Laboratory of Particle Astrophysics, Institute of High Energy Physics, Chinese Academy of Sciences, 19B Yuquan Road, Beijing, 100049, People’s Republic of China.}
\emailAdd{debabrat.d@iopb.res.in}
\emailAdd{jeetamritboudh@gmail.com}
\abstract{
The formulation of quarkyonic matter consists of treating both quarks and nucleons as quasi-particles, where a cross-over transition occurs between the two phases. This work is based upon some of the early ideas of quark matter. It has satisfied the different observational constraints on the neutron star (NS), such as its maximum mass and the canonical radius. In addition, we put an extra component inside the NS, known as Dark Matter (DM) because it is trapped due to its immense gravitational potential. In this work, we explore the impact of fermionic DM on the structure of the NS. The equation of state (EOS) is derived for the NS with the quarkyonic matter by assuming that nucleons and quarks are in equilibrium, followed by the relativistic mean-field (RMF) formalism. The recently modeled two parameterizations, such as G3 and IOPB-I, are taken to calculate the various macroscopic properties of the NS. The three unknown parameters such as the transition density ($n_t$), the QCD confinement scale ($\Lambda_{\rm cs}$), and the DM Fermi momentum ($k_f^{\rm DM}$) are varied to obtain the NS properties. The quarkyonic matter stiffens the EOS while DM softens it. The mutual combination provides good theoretical predictions for the magnitude of macroscopic properties consistent with the different observational results. Also, one can estimate the parameters of the DM admixed quarkyonic star with different statistical analyses, which can be further used to explore the other properties of the quarkyonic star.
}
\begin{document}
\maketitle
\flushbottom
\section{Introduction}
\label{intro}
A neutron star (NS) is one of the densest states of matter in the observable universe, where the central density can be as high as $5-10$ times the nuclear matter saturation density $n_0 \approx0.16$ fm$^{-3}$. Typically, it is a star having a mass $\sim$ 2 $M_\odot$ and a radius of $\sim$ 12 km, which is one of the remnants of a supernova explosion. The NSs are considered relativistic compact objects because the theory of general relativity plays a crucial role in determining their internal structure and behavior. The NSs primarily consist of densely packed neutrons, with a small fraction of protons and leptons (such as electrons and muons). However, due to their extreme densities, exotic constituents, like strangeness-bearing baryons, condensed mesons (such as pions or kaons), or even deconfined quarks may also be present. As a result, they provide valuable opportunities for the search of dense matter \cite{doi:10.1126/science.1090720, Burrows2000}. The theoretical study of the structure of NSs is crucial, because the new data on masses and radii are available, which provides effective constraints on the equation of state (EOS) of neutron star matter. Significant progress has been made both in observations and theoretical investigations in the past few years. Recently, the measurement of NS masses greater than or equal to $2 \ M_\odot$ \cite{doi:10.1146/annurev-nucl-102711-095018,2016ARA&A..54..401O} has again challenged our understanding of dense matter neutron star. The gravitational wave measurement GW170817 of binary NS merger \cite{PhysRevLett.119.161101, PhysRevLett.121.091102, PhysRevLett.121.161101, Capano2020} and NICER X-ray observations of PSR J0030+0451 \cite{Riley_2019, Miller_2019} put forward the constraint on an NS of mass $1.4 \ M_\odot$ has radius $R_{1.4}\leq 13.5$ km. To explain this recent observational constraint for NSs, Mclerran and Reddy \cite{PhysRevLett.122.122701} proposed the quarkyonic matter model where both nucleons and quark matter (QM) arise as quasi-particles due to the cross-over transition between hadrons and quarks at the core of hybrid stars where density reaches a value several times higher than nuclear saturation and preserves some of the aspects of early QM models. 
 Unlike the traditional approach of introducing QM through a first-order phase transition, such as the MIT bag model \cite{BURGIO200219} and Nambu-Jona-Lasinio model \cite{PhysRevC.60.025801}, they propose an alternative approach. In this approach, quarks drip out of nucleons and occupy the lower Fermi momentum states, resulting in a rapid increase in pressure. This is reflected in the non-monotonic rise in the speed of sound at intermediate densities followed by a decline at large densities respecting the asymptotic behavior. This model, composed of neutrons and two types of quarks ($u$ and $d$), did not initially consider the specific conditions of stellar matter. Later on, Jhao and Lattimer \cite{PhysRevD.102.023021} extended the model to include considerations such as charge neutrality, beta equilibrium between leptons and hadrons, and chemical equilibrium among quark and nucleons within the framework of quarkyonic matter. In addition to the quark species, they also accounted for two distinct nucleon species and other types of leptons in their comprehensive approach. 

Several studies are reported based on the DM capture inside the neutron star and change their macroscopic properties, as found in Refs. \cite{DM5, DM6, DM7, DM8, DM9, DM10, DM11, DM12, DM13, routray}. Till now, there is not yet a conclusive understanding of the nature of the DM particles, and numerous candidates exist, such as bosonic DM, axions, sterile neutrinos, weakly interacting massive particles (WIMPs), feebly interacting massive particles, neutralino, etc. \cite{PhysRevD.83.083512, DM2, 2017IJMPA..3230023B, Hall2010, PhysRevD.69.035001,2014JHEP...08..093H, PhysRevD.99.043016, Duffy_2009}. Various studies have suggested that the accumulation of DM can occur via scattering. The mechanism for such accumulation has been studied in Refs. \cite{dark_matter_3, dark_matter_4}. In the experimental front, many direct \cite{Bernabei2008, Bernabei2010, PhysRevLett.101.091301, doi:10.1126/science.1186112} and indirect detection methodologies are already adopted to find the evidence of DM \cite{conrad2014indirect}. In the present study, we explore the inclusion of dark matter (DM) as an additional degree of freedom within the model proposed by Jhao and Lattimer \cite{PhysRevD.102.023021}. Given the conducive environment of neutron stars (NSs) for capturing DM particles, it is plausible that DM could indeed be present within NSs, potentially exerting influence on their macroscopic properties.

The fundamental quantity that describes the dense matter systems, such as the interior of the NSs, is known as the equation of state (EOS). To determine the EOS, several microscopic non-relativistic and relativistic formalisms have been developed \cite{doi:10.1080/14786435608238186, SKYRME1958615, PhysRevC.5.626, CHABANAT1998231, PhysRevC.58.220, STONE2007587, PhysRevC.85.035201, PhysRevC.21.1568}.
One of such formalisms known as effective relativistic mean-field (E-RMF) theory describes the ground state properties of finite nuclei over the entire range of the periodic table. Its domain of applicability ranges from finite nuclei, nuclear matter,  and  NS matter \cite{PhysRevC.63.044303, PhysRevC.97.045806, patt1, patt2, patt3, patt4}. It is an effective relativistic quantum mean-field model for the nuclear many-body problem, which is fully covariant in its structure. The framework also follows the principle of self-consistency in order to derive the necessary equation of motion for nucleons whose interaction are governed by mesonic degrees of freedom \cite{WALECKA1974491, BOGUTA1977413, Serot1992, SEROT1979146}. The E-RMF model is a useful tool for understanding the behavior of dense nuclear matter at extreme concentrations because of its effectiveness and adaptability. Here, we consider the interaction between fermionic DM and nucleonic matter through the Higgs portal mechanism. We investigate NS properties by varying parameters related to both DM and QM while  utilizing the E-RMF formalism with appropriate nuclear model parameters to examine the influence of DM on quarkyonic neutron stars
\cite{DM1, DM10, PhysRevD.102.023021}.

The paper is structured as follows: Sec. \ref{Theoretical framework} provides a detailed exposition of the theoretical framework encompassing E-RMF theory, QM and DM. Following a concise overview of these models in Sec. \ref{Theoretical framework},  Sec. \ref{RD} is dedicated to deduce the macroscopic characteristics of NSs utilizing the established EOSs. Additionally, this section extensively examines the influence of DM on the NS's macroscopic attributes, offering comparative analyses with diverse observational data. At last, the result concludes in Sec. \ref{Conclusions}.

\section{Theoretical framework}
\label{Theoretical framework}
\subsection{Nuclear model}
\label{RMF}
The E-RMF formalism has demonstrated remarkable robustness in its ability to accurately describe both finite nuclei and infinite nuclear matter. The Lagrangian density is modeled by considering the interactions between different mesons and nucleons, including the self and cross-coupling among them. The parameters of the E-RMF model are determined by fitting with different experimental and empirical data. In literature, more than 200 parameters have been developed to reproduce the different experimental/observational data\cite{PhysRevC.55.540,PhysRevC.63.044303,PhysRevC.74.045806,PhysRevC.70.058801,LALAZISSIS200936,PhysRevC.82.055803,PhysRevC.82.025203,PhysRevC.84.054309,PhysRevC.85.024302,PhysRevC.97.045806,PhysRevC.102.065805}. In this E-RMF model \cite{E-RMF1, E-RMF2, E-RMF3, E-RMF4, E-RMF5}, the Lagrangian density has the self and cross-coupling between mesons up to fourth order. For completeness, the E-RMF Lagrangian for nucleon-meson-leptons for baryonic matter system is given as \citep{Reinhard_1989, KUMAR2017197, PhysRevC.97.045806}: 
\begin{eqnarray}
\label{Lagrangian}
{\cal L_{BM}} & = &  \sum_{i=n,p} \bar\psi_{i}
\Bigg\{\gamma_{\nu}(i\partial^{\nu}-g_{\omega}\omega^{\nu}-\frac{1}{2}g_{\rho}\vec{\tau}_{i}\!\cdot\!\vec{\rho}^{\,\nu})
-(M_{nucl.}-g_{\sigma}\sigma
\nonumber\\
&&
-g_{\delta}\vec{\tau}_{i}\!\cdot\!\vec{\delta})\Bigg\} \psi_{i}
-\frac{1}{2}m_{\sigma}^{2}\sigma^2
+\frac{1}{2}\partial^{\nu}\sigma\,\partial_{\nu}\sigma
+\frac{1}{2}m_{\omega}^{2}\omega^{\nu}\omega_{\nu}
\nonumber \\
&& 
-\frac{1}{4}F^{\alpha\beta}F_{\alpha\beta}
+\frac{1}{2}m_{\rho}^{2}\rho^{\nu}\!\cdot\!\rho_{\nu} 
-\frac{1}{4}\vec R^{\alpha\beta}\!\cdot\!\vec R_{\alpha\beta}
-\frac{1}{2}m_{\delta}^{2}\vec\delta^{\,2}
\nonumber \\
&&
+\frac{1}{2}\partial^{\nu}\vec\delta\,\partial_{\nu}\vec\delta
-g_{\sigma}\frac{m_{\sigma}^2}{M}\Bigg(\frac{\kappa_3}{3!}+\frac{\kappa_4}{4!}\frac{g_{\sigma}}{M}\sigma\Bigg)\sigma^3
\nonumber\\
 &&
+\frac{1}{2}\frac{g_{\sigma}\sigma}{M}\Bigg(\eta_1+\frac{\eta_2}{2} \frac{g_{\sigma}\sigma}{M}\Bigg)m_\omega^2\omega^{\nu}\omega_{\nu}
+\frac{\zeta_0}{4!}g_\omega^2(\omega^{\nu}\omega_{\nu})^2 
\nonumber\\
 &&
+\frac{1}{2}\eta_{\rho}\frac{m_{\rho}^2}{M}g_{\sigma}\sigma(\vec\rho^{\,\nu}\!\cdot\!\vec\rho_{\nu})
-\Lambda_{\omega}g_{\omega}^2g_{\rho}^2(\omega^{\nu}\omega_{\nu})(\vec\rho^{\,\nu}\!\cdot\!\vec\rho_{\nu})
\nonumber\\
 &&
+\sum_{j=e^{-},\mu} \bar\phi_{j}(i\gamma_{\nu}\partial^{\nu}-m_{j})\phi_{j}.
\label{eq1}
\end{eqnarray}
The nucleon wave functions, denoted by $\psi_{i}$, represent both neutrons and protons, while the last term in the expression accounts for the non-interacting leptonic part, namely electrons and muons. The nucleon mass is denoted by $M_{nucl.}$ (approximately 939 MeV), and specific masses and coupling constants are assigned to various mesons, including the sigma ($m_\sigma$, $g_\sigma$, $\kappa_3$, $\kappa_4$), omega ($m_\omega$, $g_\omega$, $\zeta_0$, $\eta_1$, $\eta_2$), rho ($m_\rho$, $g_\rho$, $\eta_\rho$, $\Lambda_\omega$), and delta ($m_\delta$, $g_\delta$) mesons. The field strength tensors, $F^{\alpha\beta}$ and $\vec R^{\alpha\beta}$, are employed for the omega and rho mesons, respectively.
To facilitate further calculations, we adopt the relativistic mean-field (RMF) approximation, where meson fields are replaced with their average values. This simplifies computations, especially in cases of uniform static matter, where spatial and temporal derivatives of mesons can be disregarded. Maintaining translational and rotational invariance, as well as isotropy of nuclear matter, it is crucial for ensuring the accuracy and consistency of calculations. In this approximation, only the time-like components of the isovector field and the isospin 3 component of the mesonic field are consideblue significant.  \cite{ERMF6, Serot1992, ERMF7, E-RMF8, E-RMF9, E-RMF5, PhysRevC.97.045806}. The values of the parameter sets used in the present work are given in Table \ref{tab: Lagrangian parameters}, and their nuclear matter properties at saturation density are listed in Table \ref{tab: saturation_properties}.  The energy density ($\mathcal{E}_{\rm BM}$) and pressure ($P_{\rm BM}$) for baryonic matter (BM) system can be computed from the Lagrangian (Eq. \ref{Lagrangian}) using the stress-energy tensor, yielding as \cite{E-RMF5}:  
\begin{eqnarray}
\label{eq:eden}
{\cal E}_{\rm BM} & = & \sum_{i=n,p} \frac{g_s}{(2\pi)^{3}}\int_{0}^{k_{f_{i}}} d^{3}k\, \sqrt{k^{2} + M_{i}^{*2}}\nonumber\\
&&
+n_{b} g_\omega\,\omega+m_{\sigma}^2{\sigma}^2\Bigg(\frac{1}{2}+\frac{\kappa_{3}}{3!}\frac{g_\sigma\sigma}{M_{\rm nucl.}}+\frac{\kappa_4}{4!}\frac{g_\sigma^2\sigma^2}{M_{\rm nucl.}^2}\Bigg)
\nonumber\\
&&
 -\frac{1}{4!}\zeta_{0}\,{g_{\omega}^2}\,\omega^4
 -\frac{1}{2}m_{\omega}^2\,\omega^2\Bigg(1+\eta_{1}\frac{g_\sigma\sigma}{M_{\rm nucl.}}+\frac{\eta_{2}}{2}\frac{g_\sigma^2\sigma^2}{M_{\rm nucl.}^2}\Bigg)
 \nonumber\\
&&
 + \frac{1}{2} n_3 \,g_\rho\,\rho
 -\frac{1}{2}\Bigg(1+\frac{\eta_{\rho}g_\sigma\sigma}{M_{\rm nucl.}}\Bigg)m_{\rho}^2
 \nonumber\\
 && 
-\Lambda_{\omega}\, g_\rho^2\, g_\omega^2\, \rho^2\, \omega^2
+\frac{1}{2}m_{\delta}^2\, \delta^{2}
\nonumber\\
 && 
+\sum_{j=e,\mu}  \frac{g_s}{(2\pi)^{3}}\int_{0}^{k_{F_{j}}} \sqrt{k^2 + m^2_{j}} \, d^{3}k,
\end{eqnarray}
and
\begin{eqnarray}
\label{eq:press}
P_{\rm BM}& = & \sum_{i=n,p} \frac{g_s}{3 (2\pi)^{3}}\int_{0}^{k_{f_{i}}} d^{3}k\, \frac{k^2}{\sqrt{k^{2} + M_{i}^{*2} }} \nonumber\\
&& - m_{\sigma}^2{\sigma}^2\Bigg(\frac{1}{2} + \frac{\kappa_{3}}{3!}\frac{g_\sigma\sigma}{M_{\rm nucl.}} + \frac{\kappa_4}{4!}\frac{g_\sigma^2\sigma^2}{M_{\rm nucl.}^2}\Bigg)+ \frac{1}{4!}\zeta_{0}\,{g_{\omega}^2}\,\omega^4 
\nonumber\\
&&
+\frac{1}{2}m_{\omega}^2\omega^2\Bigg(1+\eta_{1}\frac{g_\sigma\sigma}{M_{\rm nucl.}}+\frac{\eta_{2}}{2}\frac{g_\sigma^2\sigma^2}{M_{\rm nucl.}^2}\Bigg)
\nonumber\\
&&
+ \frac{1}{2}\Bigg(1+\frac{\eta_{\rho}g_\sigma\sigma}{M_{\rm nucl.}}\Bigg)m_{\rho}^2\,\rho^{2}-\frac{1}{2}m_{\delta}^2\, \delta^{2}+\Lambda_{\omega} g_\rho^2 g_\omega^2 \rho^2 \omega^2
\nonumber\\
&&
+\sum_{j=e,\mu}  \frac{g_s}{3(2\pi)^{3}}\int_{0}^{k_{F_{j}}} \frac{k^2}{\sqrt{k^2 + m^2_{j}}} \, d^{3}k,
\end{eqnarray}
where $g_s$ = 2 is the spin degeneracy of nucleons. The effective mass of nucleon (i= n, p) is defined as: 
\begin{eqnarray}
M_{i}^\star &=& M_{\rm nucl.}+ g_\sigma \sigma_0 \pm g_\delta \delta_0 
\label{eq:effm_nucleon}
\end{eqnarray}

 The momentum of a nucleon is denoted by $k_i$, while the baryonic density and isovector density are represented by $n_b$ and $n_3$ respectively. Additionally, $\sigma_0$ and $\delta_0$ stand for the mean meson fields of sigma and delta mesons respectively.

\subsection{Quarkyonic Model}
\label{QM}
Now, we discuss the quarkyonic model as proposed by Mclerran and Reddy \cite{PhysRevLett.122.122701}. In this phenomenological model, it is suggested that at the core of the NS, where the density is several times higher than the nuclear saturation density, the nucleons break into quarks. The defining characteristic of their model is its capacity to influence the correlation between mass and radius, specifically predicting a higher mass due to the stiffening of EOS as compared to baryonic. The onset of the quarkyonic phase from baryonic is marked by a swift rise in pressure, which is due to the occupation of quarks in lower momentum states as the baryon density reaches a critical value, termed transition density. This results in treating the low momentum degrees of freedom as quarks while high momenta degrees of freedom near the Fermi surface as nucleons. The momentum states near the Fermi surfaces are in order of QCD confinement scale $\Lambda_{\rm cs}$ and hence lead to the formation of nucleons, which are the bound states of quarks.

The schematic model of Mclerran and Reddy \cite{PhysRevLett.122.122701} is further developed and modified by Jhao and Lattimer \cite{PhysRevD.102.023021}. They incorporated the beta-equilibrium condition for quarkyonic matter along with charge neutrality for neutron star matter (NSM). The nucleons interact via density-dependent potential, which is fitted to select properties of uniform NM. In addition, they established chemical equilibrium among nucleons and quarks in order to establish the relation between nucleon momenta $k_{f_{(n,p)}}$ and quark momenta $k_{f_{(u,d)}}$, which marks the distinctiveness of the modified quarkyonic model. Since nucleons occupy a Fermi shell in the quarkyonic matter model, it has a finite minimum Fermi momentum $k_{0_{(n,p)}} $  and upper Fermi momentum $k_{f_{(n,p)}}$. The $d$ and $u$ quark Fermi momentum are  $k_{f_d}$ and $k_{f_u}$ respectively. The conservation law of baryon density implies the relation in \cite{PhysRevD.102.023021};
\begin{eqnarray}
n &=& n_n + n_p + \frac{n_u+n_d}{3}
\nonumber\\
&=& \frac{g_s}{6\pi^2}\bigg[(k_{f_{n}}^3-k_{0_{n}}^3)+(k_{f_{p}}^3-k_{0_{p}}^3)+\frac{(k_{f_{u}}^3+k_{f_{d}}^3)}{3}\bigg].  
\end{eqnarray}
Here, $n_n$, $n_p$, $n_u$, $n_d$  are the neutron, proton, up and down quark baryon densities.
Also, the charge neutrality condition among nucleons, quarks, and leptons leads to;
\begin{eqnarray}
n_p + \frac{2n_{u}}{3} -  \frac{n_{d}}{3}& = & n_{e^{-}} + n_{\mu}.  
\end{eqnarray}
Here, $n_{e^{-}}$  and $n_{\mu}$ are electron and muon densities respectively.

The $k_{f0_{(n,p)}}$  is related to transition Fermi momentum $k_{t_{(n,p)}}$  corresponding to transition density from nucleonic to quarkyonic matter $n_t$ by the following expression \cite{PhysRevD.102.023021}; 
\begin{eqnarray}
 k_{0(n,p)} &=& (k_{f_{(n,p)}}-k_{t_{(n,p)}})\bigg[1+ \frac{\Lambda^2}{k_{f_{(n,p)}}k_{t_{(n,p)}}}\bigg].
\end{eqnarray}
Here, 
Strong interaction equilibrium ensures that at each fixed baryon density, the Fermi gas has its lowest possible energy. It is equivalent to the condition of the chemical equilibrium among nucleons and quarks given by the relations \cite{PhysRevD.102.023021}; 
\begin{eqnarray}\label{ebnq}
\mu_n &=& \mu_u+2\mu_d,
\end{eqnarray}
\begin{eqnarray}
\mu_p &=& 2\mu_u+\mu_d,
\end{eqnarray}
where $\mu_n$, $\mu_p$, $\mu_u$, and $\mu_d$ are the chemical potentials of neutron, proton, up quark, and down quark respectively.

In addition to strong interaction equilibrium, the energy of the Fermi gas is further minimized due to the beta-equilibrium condition under the constraint of charge neutrality. This results in establishing chemical equilibrium between neutron, proton, electron, and muon, as indicated\cite{Glendenning,PhysRevD.102.023021};
\begin{eqnarray} \label{qnbe}
\mu_{n} &=& \mu_{p}+\mu_{e^{-}} , \nonumber \\
\mu_{\mu} &=& \mu_{e^{-}}.
\end{eqnarray}
Here, $\mu_{\mu}$, $\mu_{e^{-}}$ are chemical potential of muon and electron respectively. 

Also, one crucial aspect of the model is the up and down quark masses, which are not independent variables but dependent on the beta equilibrium condition of NS matter  at transition density $n_t$ via;
\begin{eqnarray}
 m_{u} = \frac{2}{3} \mu_{t_{p}} -  \frac{1}{3} \mu_{t_{n}}, \,\,\,\,\,\,\,\,\,\,\,\, m_{d} = \frac{2}{3} \mu_{t_{n}} -  \frac{1}{3} \mu_{t_{p}},
\end{eqnarray}
where $\mu_{t_{n}}$ and $\mu_{t_{p}}$ are the chemical potential of neutron and proton at $n_t$.

Since the quarks are considered as non-interacting fermion gas, their energy density ${\cal E}_{\rm QM}$ and pressure $P_{\rm QM}$ can be written as \cite{PhysRevD.102.023021};

\begin{eqnarray}
{\cal E}_{\rm QM}&=& \sum_{j=u,d}\frac{g_s N_c}{(2\pi)^3}\int_0^{k_{f_{j}}}k^2\sqrt{k^2 + m_{j}^2 }\, d^3k,
\end{eqnarray}

\begin{eqnarray}
P_{\rm QM} &=& \mu_{u} n_{u} + \mu_{d} n_{d} - {\cal E}_{\rm QM} \, .
\end{eqnarray} 

where $N_{c}$ = 3 is the color degeneracy of quarks, $m_j$ is the mass of the quark (j = u, d) and $k_j$ is the momenta of quarks.

\subsection{Dark Matter Model}
\label{DM}
In this, we choose a simple DM model, where DM particles interact with nucleons and quarks by exchanging Higgs. The Lagrangian density for this interaction is given by \cite{DM1,DM2,DM3};
\begin{eqnarray}
{\cal{L}}_{\rm DM} & = & \bar \chi \left[ i \gamma^\mu \partial_\mu - M_\chi + y h \right] \chi +  \frac{1}{2}\partial_\mu h \partial^\mu h  \nonumber\\
 & &
 - \frac{1}{2} M_h^2 h^2 + f \frac{M_{\rm nucl.}}{v} \bar \psi h \psi , 
\label{eq:LDM}
\end{eqnarray}
where $\chi$ and $\psi$ are the wave functions for the DM and nucleons, respectively and h is the Higgs field. The interaction between Higgs and nucleon is Yukawa type, having its coupling constant $f$. The $f$ is the proton-Higgs form factor. Here, we consider that the Neutralino is a DM particle having mass $M_\chi$= 200 GeV. The magnitude for the $y$ and $f$ are taken as 0.07 and 0.35, respectively, and it has been constrained using different experimental/empirical data \cite{DM5}. The Higgs mass ($M_h$) and its vacuum value ($v$) are 125 and 246 GeV, respectively. 

The energy density and pressure for the DM can be calculated with the mean-field approximation, which is given by \cite{DM1, DM3, DM2, mnras};
\begin{eqnarray}
{\cal{E}}_{\rm DM}& = & \frac{g_{s}}{(2\pi)^{3}}\int_0^{k_f^{\rm DM}} d^{3}k \sqrt{k^2 + (M_\chi^\star)^2} + \frac{1}{2}M_h^2 h_0^2 ,
\label{eq:EDM}
\end{eqnarray}
\begin{eqnarray}
P_{\rm DM}& = &\frac{g_{s}}{3(2\pi)^{3}}\int_0^{k_f^{\rm DM}} \frac{d^{3}k k^2} {\sqrt{k^2 + (M_\chi^\star)^2}} - \frac{1}{2}M_h^2 h_0^2 ,
\label{eq:PDM}
\end{eqnarray}
where $k_f^{\rm DM}$  is the DM Fermi momentum and  $M_\chi^\star$ represent DM effective mass defined in (\ref{eq:effm_tot}). 

By assuming that the nucleon density is $10^3$ times the average DM density. which implies that the mass ratio as given by $M_{\chi}/M_{\rm nucl.}=1/6$ . From this assumption, one can easily obtain the value of $k_f^{\rm DM} \approx 0.03$ GeV. The modified effective masses of nucleon ($M_{i}^\star, i = n, p$) due to their interaction with the Higgs field along with the DM effective mass ( $M_\chi^\star$) are written as; 

\begin{eqnarray}
M_{i}^\star &=& M_{\rm nucl.}+ g_\sigma \sigma_0 \pm g_\delta \delta_0 - \frac{f M_{\rm nucl.}}{v}h_0, 
\nonumber\\
M_\chi^\star &=& M_\chi - y h_0,
\label{eq:effm_tot}
\end{eqnarray}
where $h_0$ is the mean Higgs field. Now the total energy and pressure of DM admixed quarkyonic star is given as;
\begin{eqnarray}
\cal{E} = {\cal{E}}_{\rm BM} + {\cal{E}}_{\rm QM} + {\cal{E}_{\rm DM}},
\label{eq:e_total}
\end{eqnarray}
\begin{eqnarray}
P= P_{\rm BM}+ P_{\rm QM} + P_{\rm DM}.
\label{eq:p_total}
\end{eqnarray}

where  ${\cal{E}}_{\rm BM}$, ${\cal{E}}_{\rm QM}$, $\cal{E}_{\rm DM}$ are the energy density for the baryonic matter, quarkyonic matter and dark matter and   $P_{\rm BM}$,  $P_{\rm QM}$,  $P_{\rm DM}$ are the corresponding pressure respectively.

\begin{figure*}
\centering
\includegraphics[width=1.0\columnwidth]{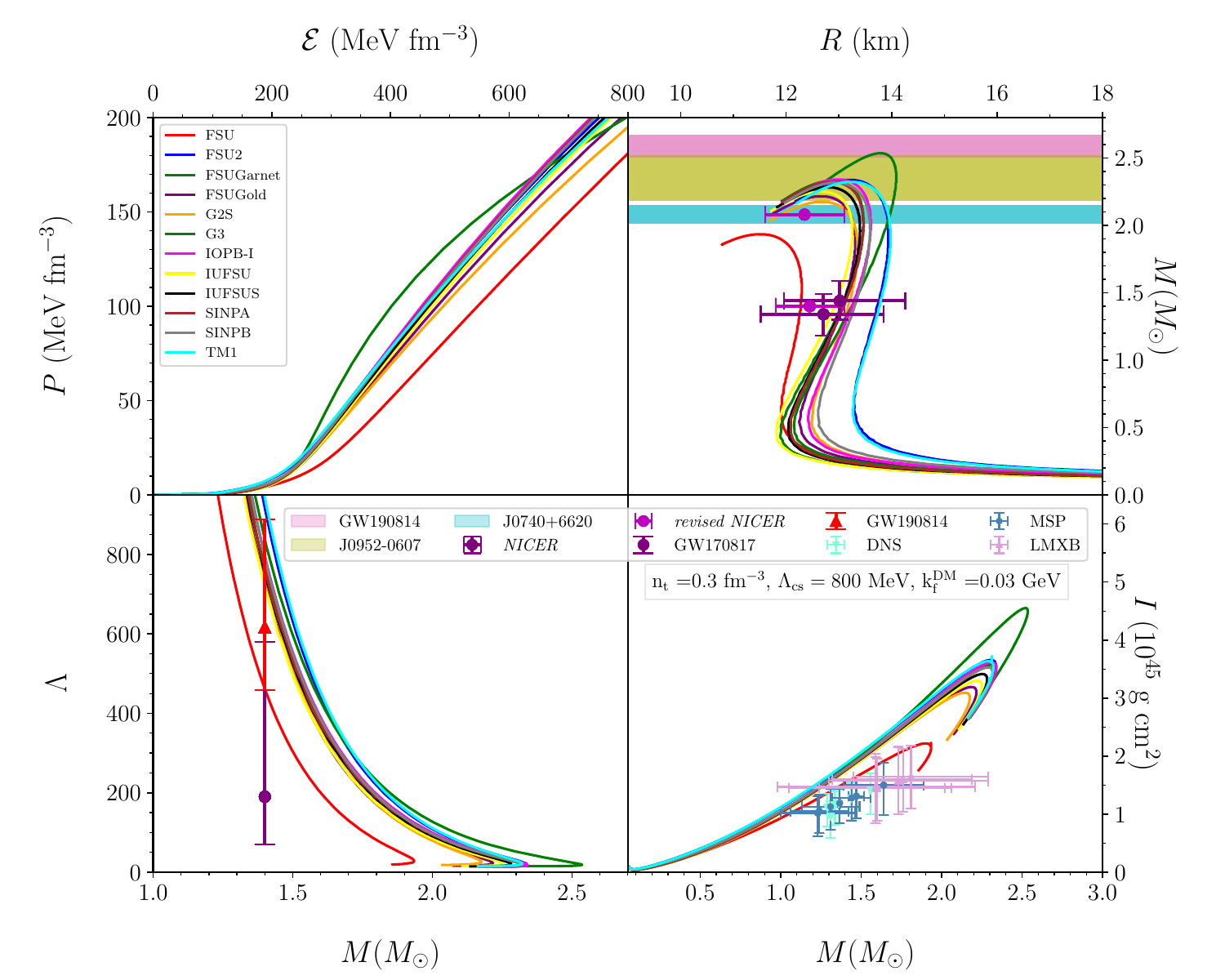}
\caption{{\it Upper left panel:} The EOSs for the 12 parameter sets with $n_t$= 0.3 $fm^{-3}$, $\Lambda_{\rm cs}$ = 800 MeV and $k_f^{\rm DM}$ =  0.03 GeV. {\it Upper right panel:} The mass-radius profile for 12 parameter sets. {\it Lower left panel:} The tidal deformability $\Lambda$ as a function of mass M. {\it Lower right panel:} The moment of inertia $I$ as a function of mass of the neutron star M.} 
\label{fig:4plot}
\end{figure*}
\begin{figure}
\centering
\includegraphics[width=0.7\columnwidth]{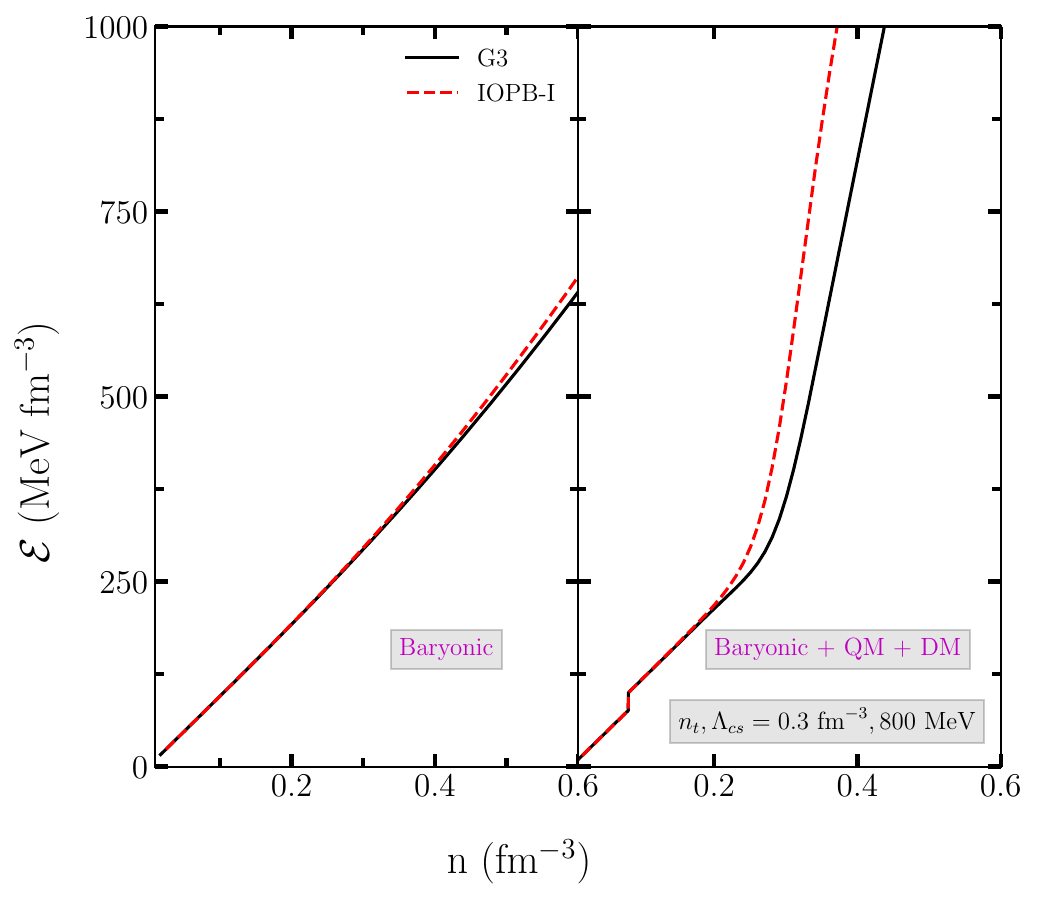}
\caption{{\it Left panel:} The EOSs ($\cal{E}$ vs $n$) for the baryonic and ({\it Right panel:)} the quarkyonic dark matter with $k_f^{\rm DM}=0.03$ GeV).  The crust-core phase transition is noticed at the total baryonic density $n=0.08$  $\rm fm^{-3}$ for G3 and IOPB-I models.}     
\label{fig:Evsrho}
\end{figure}
\begin{figure*}
\centering
\includegraphics[width=0.5\columnwidth]{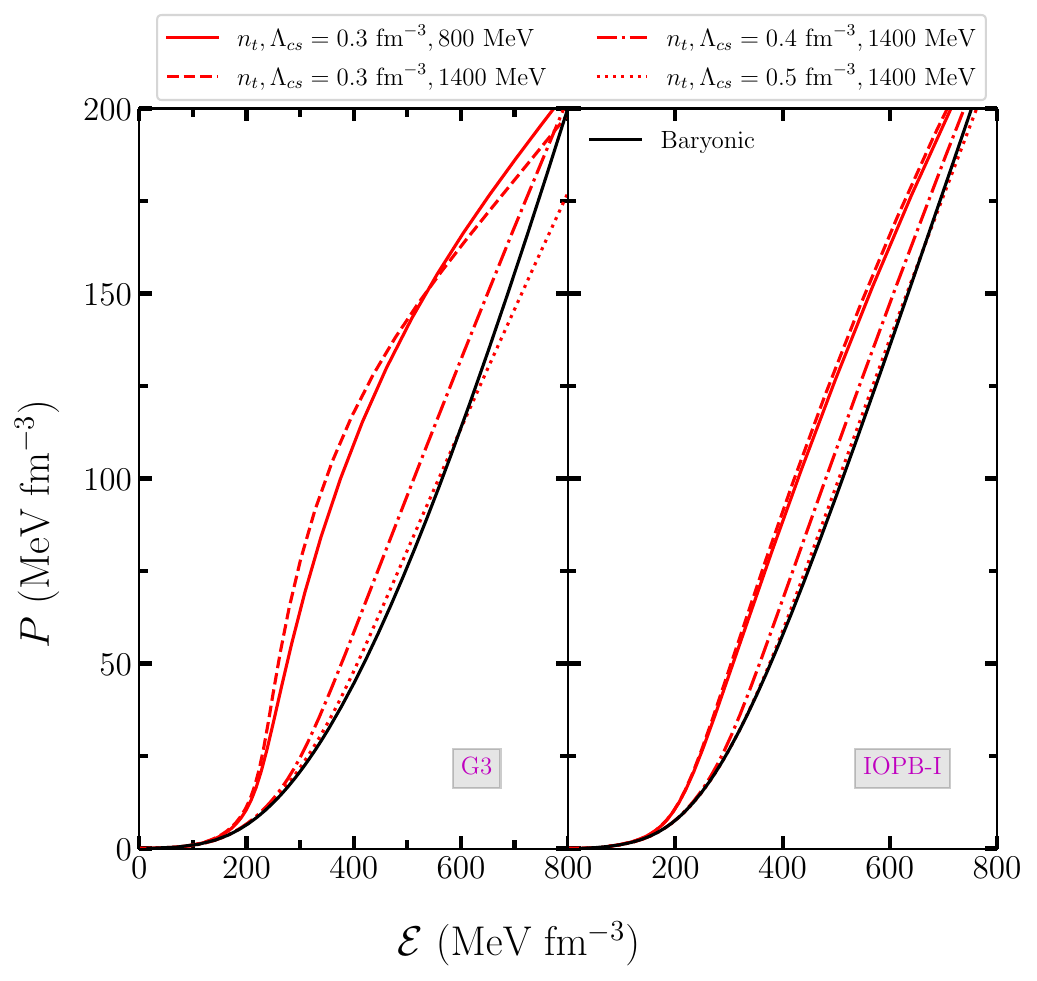}
\includegraphics[width=0.5\columnwidth]{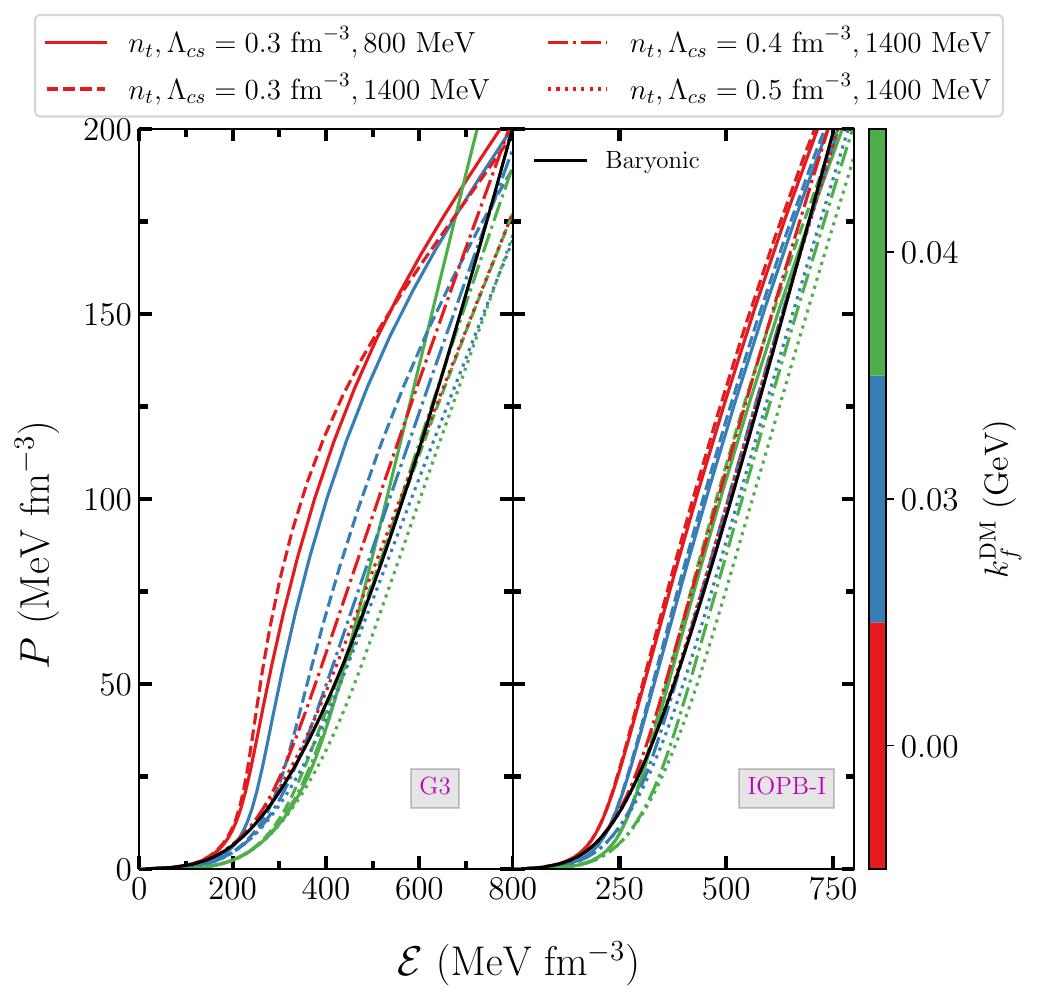}
\caption{{\it Left panel:} The EOSs for the quarkyonic matter without DM for different values of $n_t$ and $\Lambda_{\rm cs}$ with G3 and IOPB-I models. {\it Right panel:} The EOSs with DM admixed quarkyonic star having momenta, $k_f^{\rm DM}=0.00, 0.03$, and $0.04$ GeV.}
\label{fig:eos}
\end{figure*}
\begin{figure}
\centering
\includegraphics[width=0.7\columnwidth]{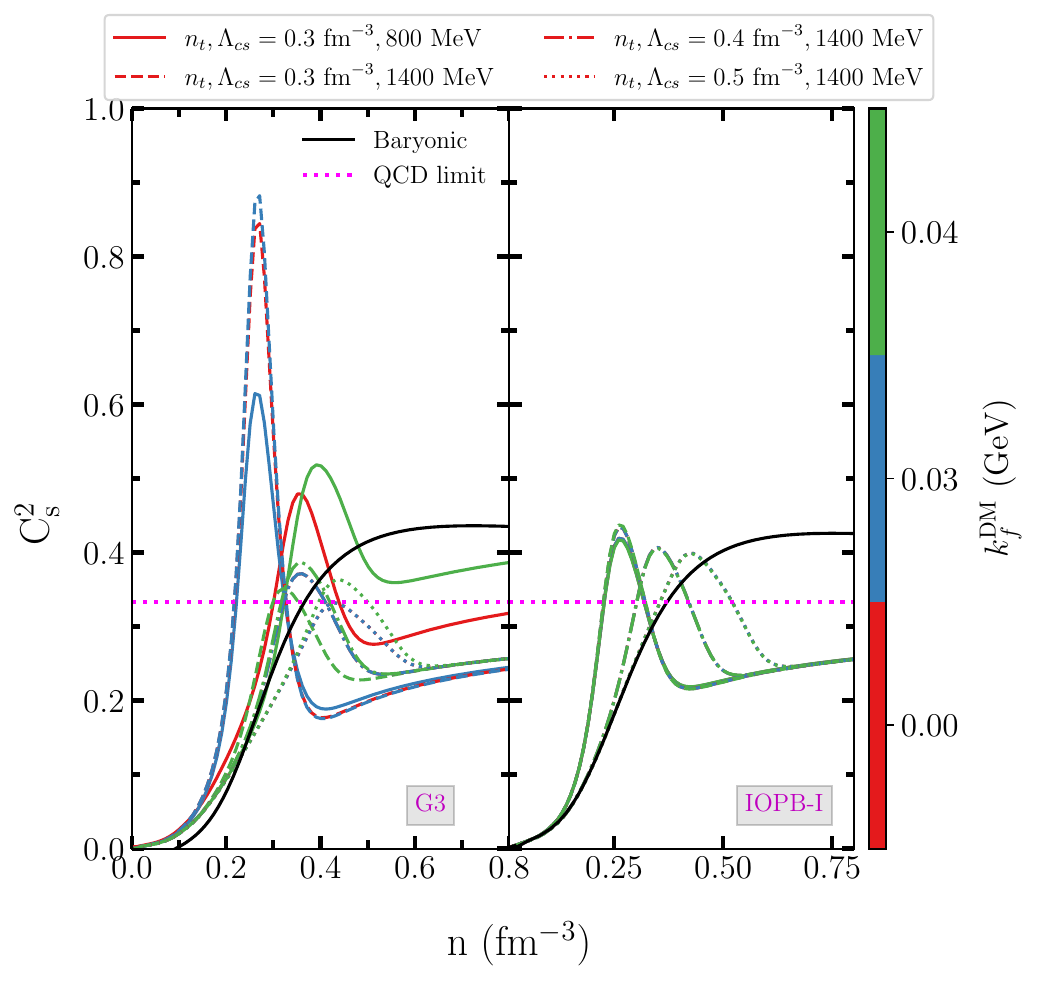}
\caption{The speed of sound $C_{\rm s}^{2}$ as a function of baryon density ($n$) for the chosen EOSs. The dotted magenta line represents the QCD conformal limit ($C_s^2 = 1/3)$.}
\label{fig:sos}
\end{figure}
\begin{figure*}
\centering
\includegraphics[width=0.5\columnwidth]{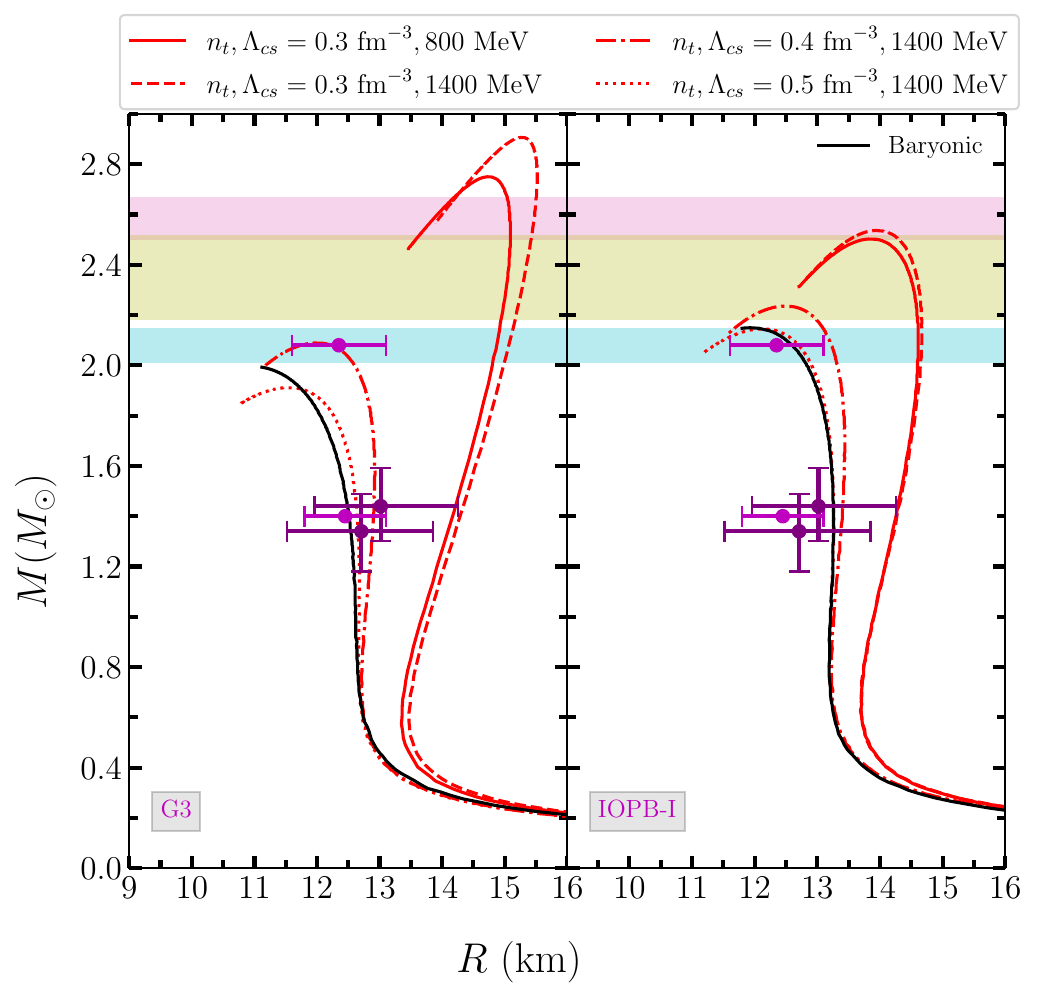}
\includegraphics[width=0.5\columnwidth]{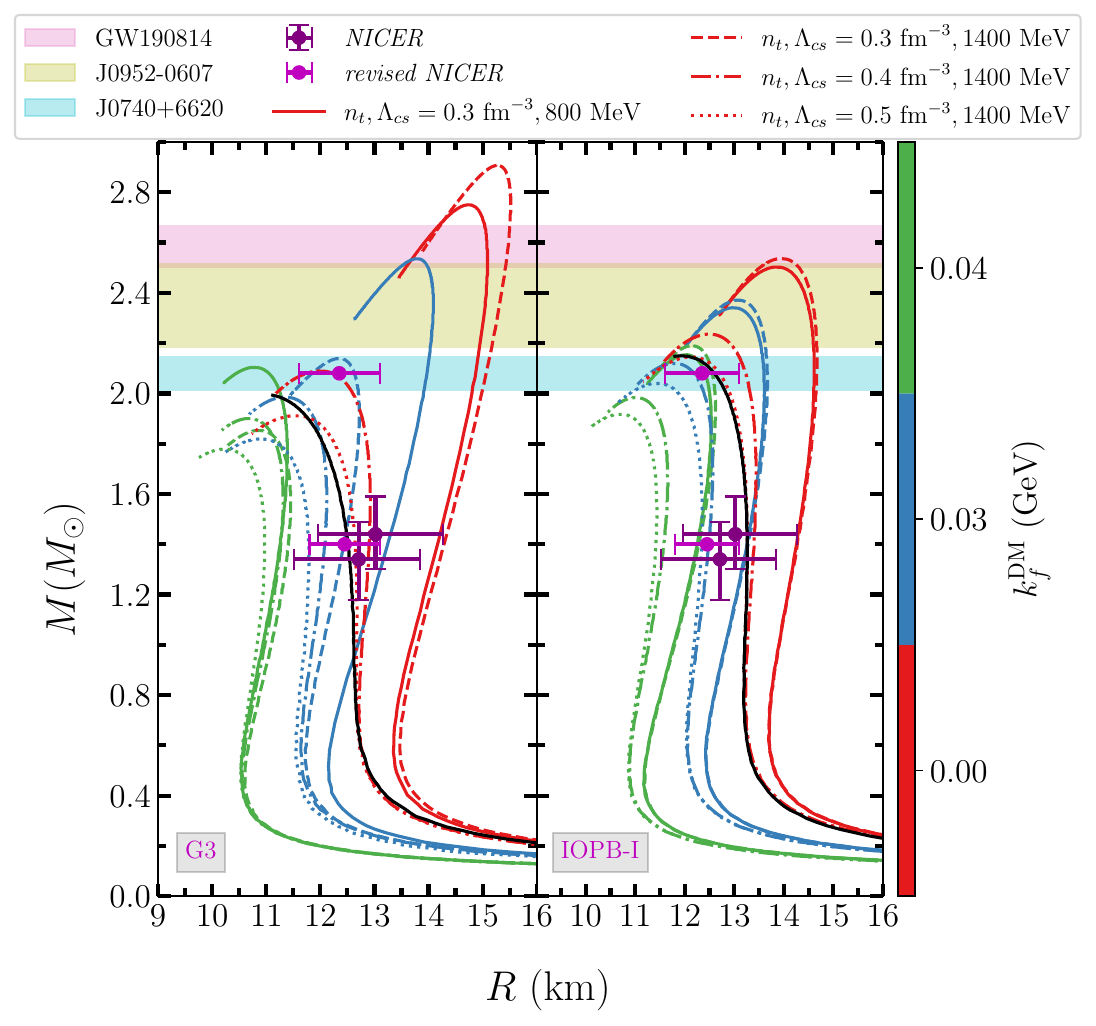}
\caption{{\it Left panel:} The $M-R$ profile without DM for different conditions of quarkyonic matter QM for G3 and IOPB-I forces. {\it Right panel:} The mass-radius diagram varies with DM momenta for different quarkyonic matter content.} 
\label{fig:mr}
\end{figure*}

\section{Results and discussions}
\label{RD}
In this section, we explain the numerical results obtained with the above formalism. The four free parameters $n_t$, $\Lambda_{\rm cs}$, $k_f^{\rm DM}$, and the model parameterizations play a crucial role in determining the behavior of NS. We describe each one in the following:

\begin{itemize}
    \item The physical significance of the parameter $n_t$ depicts the behavior of the quark's appearance at that density. The values of $n_t$  can be taken as  $0.3, 0.4$, and $0.5$ fm$^{-3}$ which are approximately $2, 2.5$, and $3$ times the nuclear saturation density ($n_0 = 0.16$ fm$^{-3}$) \cite{ankit_arxiv,PhysRevD.102.023021}. At such ultra-high density, the nucleons will overlap with each other, and quarks will drip out from the nucleon to form a fermi shell of nucleons of high momenta and quark fermi sphere of low momenta.\cite{PhysRevD.102.023021,ankit_arxiv}.

    \item The parameter $\Lambda_{\rm cs}$ is the QCD confinement scale, which acts as the cut-off momentum scale between nucleons and quarks. The low momentum degrees of freedom i.e. the momentum less than  $\Lambda_{\rm cs}$ are treated as quarks while higher momenta are treated as nucleons. In this context, the values for $\Lambda_{\rm cs}$ are 800 MeV and 1400 MeV \cite{PhysRevD.102.023021}. In the literature $\Lambda_{\rm cs}$ typically exceeds 200 MeV, marking the threshold for quark deconfinement \cite{ankit_arxiv, PhysRevLett.122.122701,Perkins}. The precise values of $\Lambda_{\rm cs}$ vary, contingent upon the theoretical framework employed for calculating the confinement scale. Consequently, these values are inherently phenomenological in nature.
    
    \item The DM Fermi momentum ($k_f^{\rm DM}$) plays a crucial role in the macroscopic properties of the NS with quarkyonic matter. One can see that $k_f^{\rm DM}$ has a direct impact on the EOS as mentioned in the Eqs. (\ref{eq:EDM}-\ref{eq:PDM}). From an agnostic assumption, we find the value around $0.03$ GeV. Therefore, we vary the limit of $k_f^{\rm DM}$ from $0.00-0.04$ GeV. In the plots, we only choose the values of $k_f^{\rm DM}=0.00, 0.03$, and $0.04$ GeV. This is due to the fact that the $k_f^{\rm DM}=0.01$, and $0.02$ GeV have no significant impact on the EOS as well as on the macroscopic properties of the NS \cite{mnras,DM1,DM2,DM3}.

    \item  In Fig \ref{fig:4plot}. {\it upper left panel} as listed in Table \ref{tab:12_NS_properties} we have shown the EOS of 12 nuclear parameter set holding other three parameter constants namely  ($n_t$, $\Lambda_{\rm cs}$, $k_f^{\rm DM} = 0.3$ fm$^{-3}$, 800 MeV, 0.03 GeV). These are RMF parameter sets which are fitted in accordance with saturation properties of nuclear matter shown in Table \ref{tab: saturation_properties}. We observe that FSU is the softest predicting mass $1.93 M_\odot$and G3 is the stiffest among all nuclear model parameter sets predicting a mass $2.54 M_\odot$ as shown in Fig \ref{fig:4plot} {\it upper right panel}. All the other 10 parameter sets predict intermediate mass values of the NS according to EOS. Then we check the sensitivity of the model parameter sets for the tidal deformability and moment of inertia as a function of mass which is tabulated in  Table \ref{tab:12_NS_properties}. Among all the 12 parameter sets we prefer G3 as one of the models for further use, because of its stiffest nature. On the other hand, We have opted for IOPB-I over FSU as a representative case of a softer EOS primarily because it successfully aligns with observational data on neutron star mass and radius. But it's important to note that many other softer EOS models could also work well. The reason is that softer EOS models can be adjusted using different quark matter parameters ($n_t$, $\Lambda_{\rm cs}$) and dark matter parameters ($k_f^{\rm DM}$) to match different observational data. Hence, onwards. we will use these two sets for the rest of our analysis.

\end{itemize}

In the next sections, we will thoroughly examine the different properties of the quarkyonic NS when mixed with DM. Additionally, we have summarized these NS properties in Table \ref{tab:all parameter set} for all combinations of free parameters.

\begin{table*}[]
\centering
\renewcommand{\tabcolsep}{0.05cm}
\renewcommand{\arraystretch}{1.6}
\caption{The coupling constants of the Lagrangian (Eq. \ref{eq1}) for the 12 parameter sets \cite{Parmar_2022}. }
\label{tab: Lagrangian parameters}
\begin{tabular}{cccccccccccccc}
\hline \hline
Model     & FSU & FSU2  & FSUGarnet & FSUGold & G2S & G3 & IOPB-I & IUFSU & IUFSUS & SINPA & SINPB & TM1 \\
\hline 
$m_{\sigma}/M$      & 0.523    &0.529   & 0.529  &  0.523   & 0.554  & 0.559        & 0.533        &  0.523    & 0.543  & 0.527    & 0.525  & 0.544 \\
$m_{\omega}/M$      & 0.833    & 0.833  & 0.833  &   0.833  & 0.833  & 0.832        &  0.833       &  0.833    & 0.833  & 0.833    & 0.833  & 0.833\\
$m_{\rho}/M$        & 0.812    & 0.812  & 0.812  &  0.812   & 0.812  & 0.820        &  0.812       &  0.812    & 0.819  & 0.812    & 0.812  & 0.820    \\
$m_{\delta}/M$      & 0.000    & 0.000  & 0.000  &   0.000  &  0.000 & 1.043        & 0.000        &  0.000    & 0.000  & 0.000    & 0.000  & 0.000  \\
$g_{\sigma}/4 \pi$  &  0.842   & 0.827  & 0.837  &  0.842   &  0.835 & 0.782        &  0.827       &  0.793    & 0.837  &-0.845    & -0.843 & 0.798 \\
$g_{\omega}/4 \pi$  & 1.138    & 1.078  & 1.091  &  1.138   &  1.016 & 0.923        & 1.062        &  1.037    & 1.066  & 1.102    & 1.104  & 1.003 \\
$g_{\rho}/4 \pi$    &  0.936   & 0.713  &  1.105 &  0.936   &  0.938 & 0.962        & 0.885        &  1.081    & 0.988  & 1.021    &0.845   & 0.368\\
$g_{\delta}/4 \pi$  &    0.000 & 0.000  &  0.000 &  0.000   &  0.000 & 0.160        & 0.000        &  0.000    & 0.000  & 0.000    & 0.000  & 0.000 \\
$k_{3} $            &  1.420   &  3.002 & 1.368  &  1.420   & 3.247  & 2.606        & 1.496        &  1.159    & 1.141  & 1.537    &  1.486 &-7.232  \\
$k_{4}$             &  0.023   &  0.000 & -1.397 &  0.023   & 0.632  & 1.694        &-2.932        &  0.096    & 1.032  &   -1.190 & -0.802 & 0.618\\
$\zeta_{0}$         &    0.060 &  0.025 &  4.410 &  0.060   & 2.642  & 1.010        & 3.103        &  0.030    & 5.389  &  5.363   &  5.467 & 71.307 \\
$\eta_{1}$          &    0.000 &  0.000 &  0.000 &  0.000   & 0.650  & 0.424        & 0.000        &  0.000    & 0.000  &   0.000  &   0.000& 0.000\\
$\eta_{2}$          &    0.000 &  0.000 & 0.000  &  0.000   & 0.110  & 0.114        & 0.000        &  0.000    & 0.000  &  0.000   &  0.000 & 0.000\\
$\eta_{\rho}$       &    0.000 &  0.000 &  0.000 &  0.000   & 4.490  & 0.645        & 0.000        &  0.000    & 0.000  &   38.179 & 13.487 & 0.000\\
$\Lambda_{\omega}$  &   0.030  &  0.000 & 0.000  &  0.030   & 0.000  & 0.038        & 0.024        &  0.046    & 0.000  & 0.000    & 0.000  & 0.000\\
\hline
\hline
\end{tabular}  
\end{table*}

\begin{table*}
\centering
\caption{The saturation properties of nuclear matter such as density ($\rho_{\rm sat}$), binding energy per particle ($B/A$), effective mass ratio ($M^*/M$), incompressibility constant ($K$), symmetry energy ($J$), slope parameter ($L$) and curvature of symmetry energy ($K_{\rm sym}$) for the 12 E-RMF parameter sets \cite{Parmar_2022}. The experimental/empirical values are also listed.}
\label{tab: saturation_properties}
\renewcommand{\tabcolsep}{0.05cm}
\renewcommand{\arraystretch}{1.5}
\begin{tabular}{llllllll}
\hline \hline
\begin{tabular}[c]{@{}l@{}} Parameter \\ sets \end{tabular}  &\hspace{0.1cm} $\rho_{\rm sat}$ &\hspace{0.1cm} $B/A$ & $M^*/M$ & \hspace{0.1cm}$K$ &\hspace{0.1cm} $J$ & \hspace{0.1cm}$L$ &\hspace{0.1cm} $K_{sym}$ \\ \hline 
FSU  & 0.148 & -16.28 & 0.61 & 229.54 & 37.42 & 109.62 & 2.64 \\ 

FSU2  & 0.150 & -16.28 & 0.59 & 238.00 & 37.62 & 112.80 & -24.25 \\ 
FSUGarnet  & 0.153 & -16.23 & 0.57 & 229.50 & 30.95 & 51.04 & 59.36 \\ 

FSUGold & 0.148 & -16.28 & 0.61 & 229.54 & 32.56 & 60.44 & -51.40 \\ 

G2S & 0.154 & -16.07 & 0.66 & 214.77 & 30.39 & 69.68 & -21.93 \\ 

G3  & 0.148 & -16.02 & 0.69 & 243.96 & 31.84 & 49.31 & -106.07 \\
IOPB-I  & 0.149 & -16.10 & 0.65 & 222.65 & 33.30 & 63.58 & -37.09 \\ 
IUFSU & 0.155 & -16.40 & 0.67 & 231.33 & 31.30 & 47.21 & 28.53 \\ 
IUFSUS  & 0.150 & -16.10 & 0.58 & 236.00 & 29.85 & 51.508 & 7.87 \\ 
SINPA  & 0.151 & -16.00 & 0.58 & 203.00 & 31.20 & 53.86 & -26.75 \\ 
SINPB  & 0.150 & -16.04 & 0.63 & 206.00 & 33.95 & 71.55 & -50.57 \\ 
TM1 & 0.145 & -16.30 & 0.63 & 281.00 & 36.94 & 111.00 & 34.00 \\ 
\hline
EMP./EXP.  & 0.148 -- 0.185    & -(15.0 -- 17.0)   & 0.55 -- 0.6 &  220 -- 260  & 33.4 -- 42.8 & 69 -- 143   & - (174.0 -- 31.0)\\
 &   \cite{saturation_density}  &    \cite{saturation_density} &  \cite{effective_mass}&   \cite{K} & \cite{patt3}&  \cite{patt3} &  \cite{Ks} \\
\hline
\hline
\end{tabular}%
\end{table*}

\begin{sidewaystable}[]
\centering
\renewcommand{\tabcolsep}{0.05cm}
\renewcommand{\arraystretch}{1.5}
\caption{The neutron star properties such as maximum mass ($M_{\rm max}$) and maximum radius ($R_{\rm max}$) for the 12 parameter sets are listed. The canonical values for radius ($R_{1.4}$), dimensionless tidal deformability ($\Lambda_{1.4}$), moment of inertia (MOI) ($I_{1.4}$) (in unit $10^{45}$ g cm$^2$) are also given at $n_t=0.3$ $^fm^{-3}$, $\Lambda_{cs}=800$ MeV and $k_f^{\rm DM}$ (GeV) = 0.03 GeV. The symbol \ding{52} satisfies the observational data and \ding{55} represents the deviation.}
\label{tab:12_NS_properties}
\begin{tabular}{cccccccccccccc}
\hline \hline
Model     & $M_{\rm max}$ & $R_{\rm max}$ & $R_{1.4}$ & $\Lambda_{1.4}$ & $I_{1.4}$ & NICER & Revised NICER & J0740+6620 & J0952-0607 & GW170817 & \multicolumn{2}{l}{GW190814} \\
          & ($M_\odot$)   & (km)          & (km)      &                 &           & (R$_{1.4}$) & (R$_{1.4}$)   & ($M_{\rm max}$) & ($M_{\rm max}$) & ($\Lambda$) & (M/$\Lambda$) \\
\hline
FSU       & 1.93          & 11.50         & 12.28     & 450.96          & 1.54      & \ding{55} & \ding{52}     & \ding{55}       & \ding{55}       & \ding{52}   & \ding{55} /\ding{55} \\
FSU2      & 2.34          & 13.21         & 13.73     & 908.38          & 1.85      & \ding{52} & \ding{55}     & \ding{55}       & \ding{52}       & \ding{55}   & \ding{55} /\ding{55} \\
FSUGarnet & 2.33          & 12.88         & 13.07     & 707.92          & 1.80      & \ding{52} & \ding{52}     & \ding{55}       & \ding{52}       & \ding{55}   & \ding{55} /\ding{52} \\
FSUGold   & 2.21          & 12.67         & 13.09     & 680.25          & 1.76      & \ding{52} & \ding{52}     & \ding{55}       & \ding{55}       & \ding{55}   & \ding{55} /\ding{52} \\
G2S       & 2.17          & 12.68         & 13.19     & 707.66          & 1.77      & \ding{52} & \ding{55}     & \ding{55}       & \ding{55}       & \ding{55}   & \ding{55} /\ding{52} \\
G3        & 2.54          & 13.80         & 13.24     & 838.3           & 1.80      & \ding{52} & \ding{55}     & \ding{55}       & \ding{55}       & \ding{55}   & \ding{52} /\ding{52} \\
IOPB-I    & 2.34          & 12.99         & 13.26     & 796.46          & 1.77      & \ding{52} & \ding{55}     & \ding{55}       & \ding{52}       & \ding{55}   & \ding{55} /\ding{52} \\
IUFSU     & 2.25          & 12.66         & 12.92     & 706.51          & 1.73      & \ding{52} & \ding{52}     & \ding{55}       & \ding{55}       & \ding{55}   & \ding{55} /\ding{52} \\
IUFSUS    & 2.28          & 12.79         & 13.08     & 706.434         & 1.78      & \ding{52} & \ding{52}     & \ding{55}       & \ding{52}       & \ding{55}   & \ding{55} /\ding{52} \\
SINPA     & 2.32          & 12.86         & 13.10     & 761.68          & 1.76      & \ding{52} & \ding{52}     & \ding{55}       & \ding{52}       & \ding{55}   & \ding{55} /\ding{52} \\
SINPB     & 2.31          & 12.93         & 13.31     & 759.8           & 1.79      & \ding{52} & \ding{55}     & \ding{55}       & \ding{52}       & \ding{55}   & \ding{55} /\ding{52} \\
TM1       & 2.32          & 13.24         & 13.78     & 924.74          & 1.87      & \ding{52} & \ding{55}     & \ding{55}       & \ding{52}       & \ding{55}   & \ding{55} /\ding{55} \\
\hline \hline
\end{tabular}
\end{sidewaystable}

\subsection{Equation of states}
In this sub-section, we obtain the EOS for DM admixed quarkyonic NS. The detailed methodology is discussed in Section \ref{Theoretical framework}. A comprehensive study regarding the EOS of quarkyonic NS within the E-RMF approach is available in Refs \cite{PhysRevD.102.023021, ankit_arxiv}. In the crustal region, we rely on the EOS established in Ref. \cite{Parmar_2022}. Typically, G3 exhibits a softer EOS compared to IOPB-I, attributed to the presence of higher-order couplings within its Lagrangian density. Consequently, the macroscopic properties of NS derived from G3 are generally smaller in magnitude than those obtained from IOPB-I. However, this scenario undergoes a significant transformation with the introduction of both DM and QM. The inclusion of quark degrees of freedom notably stiffens the EOS yielding a pronounced increase in the speed of sound at intermediate density, a crucial feature for reconciling with observable constraints. This heightened speed of sound necessitates an asymptotic trend, reaching a value $(1/\sqrt{3})$ and leading to the construction and examination of the parametrized speed of sound models \cite{sos1}. However, the quarkyonic model presents an alternative, physically intuitive theoretical framework with higher mass prediction than baryonic scenarios. The incorporation of dark matter exerts an opposite effect. Predictions for mass and radius are diminished relative to their purely baryonic counterparts. Therefore, the presence of a stiff EOS in the quarkyonic model allows us to study the potential dark matter influences and constraints of these two components inside the NS in accordance with observational data.

In Fig. \ref{fig:eos}, a comparison of EOS ($\cal{E}$ vs $n$) is drawn between the baryonic case (left panel) and quarkyonic dark matter admixed neutron star (right panel). As mentioned, for this purpose, we have selected two parameter sets, G3 and IOPB-I. In the latter scenario, we have considered a particular combination of the other three parameters ($n_t$, $\Lambda_{\rm cs}$, $k_f^{\rm DM}$ = $0.3$ fm$^{-3}$, 800 MeV, 0.03 GeV) as an example. In the baryonic case, we notice the IOPB-I set is stiffer than the G3 parameter (see also Fig. \ref{fig:eos}). On the contrary, the addition of quark matter stiffens the EOS in both G3 and IOPB-I. A further inspection of the EOSs of baryonic and dark matter admixed cases reveals that the rate of change in EOSs is more prominent in G3 than the IOPB-I, which is reflected in the larger values of speed of sound $C_s^2$ and mass of the NSs for the G3 set (Figs. \ref{fig:sos} and \ref{fig:mr}). We depict the EOSs for various combinations of $n_t$, $\Lambda_{\rm cs}$, and $k_f^{\rm DM}$ in Fig. \ref{fig:eos}. In literature, the variation of $n_t$ for different properties of the quarkyonic NS has already been discussed \cite{ankit_arxiv}. However, our primary motivation is to see the effect of DM in this study.  The different combinations predict the EOSs and macroscopic properties of the star which can be fixed/constrained using the observational data. 

The EOSs in the left panel for quarkyonic NS without DM and in the right panel with DM having Fermi momenta $0.00$, $0.03$, and $0.04$ GeV are depicted in Fig. \ref{fig:eos}. From the left panel of the figure, it is observed that the EOSs for each scenario get stiffened due to the addition of QM. This is due to the quark interactions, which increase the magnitude of both the energy density and pressure of the system compared to the baryonic case (black line).In the case of the DM admixed quarkyonic NS, the system becomes more complex, which is evident from the right panel of the figure. With the addition of DM, the system loses energy. Therefore, the pressure of the system decreases, which softens the EOS. An increased amount of DM softens the EOS more. Therefore, EOSs with $0.04$ GeV are the softer ones than $0.00$ and $0.03$ GeV. In some cases, the EOSs are almost the same as the baryonic case. Therefore, we observe that the QM stiffens the EOS; however, we found DM does the reverse. This type of scenario is beneficial to explain the different observational constraints of NS, which will be discussed in the following sub-sections. However, one crucial analysis we want to check is the causality for the DM admixed quarkyonic star.

\subsection{Causality Test}
 
Now, we calculate the speed of sound to test the causality conditions for all the cases of EOS. Although, the causality condition is already tested for the quarkyonic star in Refs. \cite{ankit_arxiv, PhysRevD.102.023021}. It is also essential, for DM admixed quarkyonic star, to satisfy; otherwise, the system is considered unphysical. Therefore, we calculate the speed of sound for all suitable conditions varying $n_t$, and $\Lambda_{\rm cs}$ with different DM momenta as shown in Fig. \ref{fig:sos}. Since the E-RMF model is already causal, the black line satisfied the causality limit. In the case for DM admixed quarkyonic star, both causality limit ($C_s^2=1$)  and  QCD conformal limit ($C_s^2=1/3$) are satisfied, except in one case in the G3 panel ($n_t$, $\Lambda_{\rm cs}$, $k_f^{\rm DM} = 0.3$ fm$^{-3}$, 800 MeV, 0.04 GeV) which violates the later condition. One should exclude such cases in the study. Therefore, with those EOSs, we can calculate the well-known observables for a star, such as mass, radius, tidal deformability, etc., in the following sub-sections which are within the causality limit.
\begin{sidewaystable}[]
\centering
\renewcommand{\tabcolsep}{0.05cm}
\renewcommand{\arraystretch}{1.2}
\caption{DM admixed quarkyonic star properties for G3 and IOPB-I parameter sets. The unit of moment of inertia (MOI) is $10^{45}$ g cm$^2$. The symbol \ding{52} satisfies the observational data and \ding{55} represents the deviation.  }
\begin{tabular}{cccccccccccccccc}
\hline \hline
Model  & $n_{\rm t}$ & $\Lambda_{\rm cs }$ & $k_f^{\rm DM}$ & $M_{\rm max}$ &  $R_{\rm max}$ & $R_{1.4}$  & $\Lambda_{1.4}$ & $I_{1.4}$ &NICER & Revised NICER & J0740+6620 & J0952-0607  & GW170817 & GW190814 \\
  & (fm$^{-3}$) & (MeV) & (GeV) & ($M_\odot$) & (km)& (km) &   &  & \hspace{0.5mm} R$_{1.4}$  & \hspace{0.5mm} R$_{1.4}$ & ($M_{\rm max}$) & ($M_{\rm max}$) & ($\Lambda$)& (M/$\Lambda$)\\
\hline
G3 & 0.00 & 0.00  & 0.00 &1.99&10.94 &12.11 & 464.63   & 1.53 & \ding{52} & \ding{52}  & \ding{55}& \ding{55}  & \ding{52} & \ding{55} / \ding{52}      \\
G3 & 0.3 & 800  & 0.00 & 2.75 &14.74 &14.17 & 1181.62   & 1.98  & \ding{55} & \ding{55}  & \ding{55}& \ding{55}  & \ding{55} & \ding{55} / \ding{55}      \\
G3 & 0.3 & 800  & 0.03 & 2.54 &13.80 &13.24 & 838.30    & 1.80  & \ding{52} & \ding{55} & \ding{55} & \ding{55} & \ding{55} & \ding{52}/\ding{52}    \\
G3 & 0.3 & 800  & 0.04 & 2.10 &10.77 &11.26 & 290.19     & 1.36  & \ding{55} & \ding{55} & \ding{52} & \ding{55} & \ding{52}  & \ding{55}/\ding{55}   \\[0.2cm] 
G3 & 0.3 & 1400 & 0.00 & 2.91 &15.29 &14.29 & 1251.80   & 2.01  & \ding{55} & \ding{55} & \ding{55}& \ding{55}& \ding{55}  & \ding{55}/\ding{55}     \\
G3 & 0.3 & 1400 & 0.03 & 2.14 &12.34 &12.44 & 535.60    & 1.59  & \ding{52} & \ding{52}   & \ding{52}  & \ding{55} & \ding{52} & \ding{55}/\ding{52}     \\
G3 & 0.3 & 1400 & 0.04 & 1.85 &10.86 &11.41  & 317.15    & 1.39  & \ding{55} & \ding{55}& \ding{55} & \ding{55} & \ding{52} & \ding{55}/\ding{55}  \\[0.2cm] 
G3 & 0.4 & 1400 & 0.00 & 2.09 &12.00 &12.90   & 599.05    & 1.64 & \ding{52} & \ding{52} & \ding{52} & \ding{55} & \ding{55}& \ding{55}/\ding{52}      \\
G3 & 0.4 & 1400 & 0.03 & 1.99 &11.35 &12.08  & 424.00    & 1.50  & \ding{52} & \ding{52} & \ding{55}& \ding{55} & \ding{52} &  \ding{55}/\ding{55}   \\
G3 & 0.4 & 1400 & 0.04 & 1.90 &10.68 &11.27  & 291.60    & 1.36 & \ding{55}  & \ding{55} & \ding{55} & \ding{55}  & \ding{52} & \ding{55}/\ding{55}   \\[0.2cm] 
G3 & 0.5 & 1400 & 0.00 & 1.91 &11.55 &12.64 & 505.60    & 1.57 & \ding{52} & \ding{52} & \ding{55} & \ding{55} & \ding{52} & \ding{55}/\ding{52}    \\
G3 & 0.5 & 1400 & 0.03 & 1.82 &10.90 &11.79 & 352.57 & 1.42 & \ding{52} & \ding{52}  & \ding{55} & \ding{55}  & \ding{52} & \ding{55}/\ding{55}  \\
G3 & 0.5 & 1400 & 0.04 & 1.78 &10.23 &10.97  & 232.43    & 1.28  & \ding{55} & \ding{55}  & \ding{55}& \ding{55}& \ding{52} &  \ding{55}/\ding{55}   \\
\hline \hline
IOPB-I & 0.00 & 0.00  & 0.00 & 2.15 &11.94 &12.78 & 689.62 & 1.70  & \ding{52}   & \ding{52}  & \ding{52}  & \ding{55}   & \ding{55} & \ding{55}/\ding{52}     \\
IOPB-I & 0.3 & 800  & 0.00 & 2.50 &13.86 &14.26 & 1159.70 & 1.97  & \ding{55}   & \ding{55}  & \ding{55}  & \ding{52}   & \ding{55} & \ding{52}/\ding{55}     \\
IOPB-I & 0.3 & 800  & 0.03 & 2.34 &12.99 &13.26 & 796.46  & 1.77 & \ding{52}  & \ding{55}   & \ding{55}   & \ding{52} & \ding{55}  & \ding{55} /\ding{52}    \\
IOPB-I & 0.3 & 800  & 0.04 & 2.15 &12.03 &12.25 & 519.72   & 1.58 & \ding{52}  & \ding{52} & \ding{52} & \ding{55}  & \ding{52}  & \ding{55} / \ding{52}    \\[0.2cm] 
IOPB-I & 0.3 & 1400 & 0.00 & 2.54 &13.94 &14.27  & 1169.31 & 1.98 & \ding{55}  & \ding{55}  & \ding{55}  & \ding{55}  & \ding{55}  & \ding{52} /\ding{55}    \\
IOPB-I & 0.3 & 1400 & 0.03 & 2.37 &13.08 &13.28  & 804.95 & 1.78 & \ding{52}  & \ding{55}   & \ding{55} & \ding{52} & \ding{55}   & \ding{55} /\ding{52}\\
IOPB-I & 0.3 & 1400 & 0.04 & 2.19 &12.15 &12.28   & 530.93 & 1.59 & \ding{52}  & \ding{52}  & \ding{55} & \ding{52} & \ding{52}   & \ding{55} / \ding{52}   \\[0.2cm] 
IOPB-I & 0.4 & 1400 & 0.00 & 2.24 &12.49 &13.40   & 743.63  & 1.74 & \ding{52}  & \ding{55}  & \ding{55}   & \ding{52} & \ding{55}  & \ding{55} /\ding{52} \\
IOPB-I & 0.4 & 1400 & 0.03 & 2.12 &11.79 &12.49  & 515.81  & 1.58 & \ding{52}  & \ding{52} & \ding{52}& \ding{55} & \ding{52}  & \ding{55} /\ding{52}\\
IOPB-I & 0.4 & 1400 & 0.04 & 1.98 &11.01 &11.58  & 342.53   & 1.42 & \ding{52}  & \ding{55}  & \ding{55} & \ding{55}   & \ding{52} & \ding{55}/\ding{55}   \\[0.2cm] 
IOPB-I & 0.5 & 1400 & 0.00 & 2.15 &12.12 &13.27 &  685.78 & 1.70 & \ding{52}  & \ding{55}  & \ding{52} & \ding{55}  & \ding{55}   & \ding{55} /\ding{52}  \\
IOPB-I & 0.5 & 1400 & 0.03 & 2.04 &11.45 &12.34  & 467.08  & 1.54 & \ding{52}  & \ding{52} & \ding{52} & \ding{55}  & \ding{52}  &  \ding{55} /\ding{52}  \\
IOPB-I & 0.5 & 1400 & 0.04 & 1.92 &10.71 &11.42 & 304.12   & 1.38 & \ding{55}  & \ding{55}  & \ding{55}  & \ding{55}  & \ding{52} &  \ding{55} /\ding{55}    \\
\hline \hline
\end{tabular}
\end{sidewaystable}
\label{tab:all parameter set}
\subsection{Mass and Radius relations}
\begin{figure*}
\includegraphics[width=1\columnwidth]{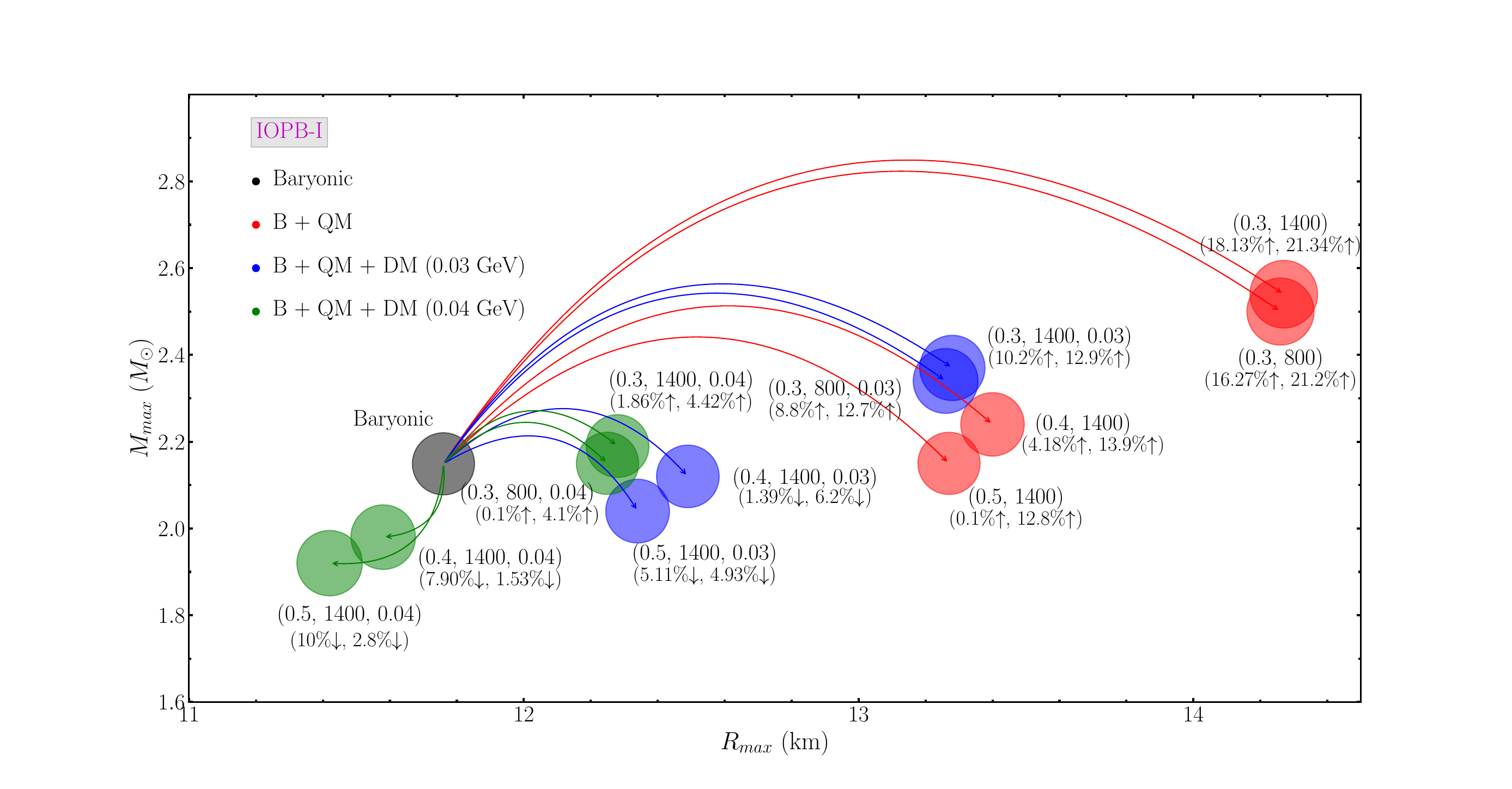}
\caption{The mass ($M_{max}$) and radius ($R_{max}$) are
plotted with IOPB-I model with different admixed of quarkyonic and dark matter contents. The center of the circle is plotted as ($R_{max}$, $M_{max}$) obtained from our calculations. The combination of quarkyonic and dark matter configurations is represented as ($n_{\rm t}$, $\Lambda_{\rm cs}$, $k^{\rm DM}_f$). The relative change in  mass and radius with respect to the pure  baryonic star is given by ($\delta M_{\rm rel}\%$, $\delta R_{rel}\%$).}
\label{fig:bubble}
\end{figure*}

The mass ($M$) and radius ($R$) for the static NS embedded within  spherically symmetric and isotropic space-time  metric are determined by a set of coupled differential equations, which are known as the TOV equations \cite{TOV1, TOV2} defined as:

\begin{eqnarray}
\frac{dp(r)}{dr} &=& - \frac{[p(r)+{\epsilon(r)}][m(r)+4\pi r^3 p(r)]}{r[r-2m(r)]}, \\ 
\frac{dm(r)}{dr} &=& 4\pi r^2 \epsilon(r),
\end{eqnarray}
\\
where ${\epsilon(r)}$, $p(r)$, and $m(r)$ are the internal energy density, pressure, and mass of the star as a function of radial coordinate respectively. These star variables are obtained using \eqref{eq:e_total} and \eqref{eq:p_total} and integrating the above equations. We solved these equations using the proper initial conditions, where the EOS is taken as input. The corresponding $M-R$ profiles are depicted in Fig. \ref{fig:mr} along with the observational constraints for different DM percentages with varying QM parameters. As already discussed in the earlier subsections, the G3 EOS became stiffer due to quarkyonic interactions. Therefore, the magnitude of $M-R$ profiles is also increased compared to the baryonic one. In the case of $n_t, \Lambda_{\rm cs} = 0.3$ fm$^{-3}$, $800 \ (1400)$ MeV, the $M-R$ relations predict the mass more than the GW190814 ($M=2.50-2.67 \ M_\odot$) case. For lower cut off values of $n_t, \Lambda_{\rm cs} = 0.5$ fm$^{-3}$, $1400$ MeV, we get lesser mass $2.0 \ M_\odot$. Therefore, without the DM case, one can exclude those extreme cases from the study. However, the scenario differs for the IOPB-I case, which satisfies almost all constraints overlaid in Fig. \ref{fig:mr}. In Table \ref{tab:all parameter set}, we provide all numbers for mass and radius corresponding to the maximum and canonical cases for all variations of $n_t$, $\Lambda_{\rm cs}$, and $k_f^{\rm DM}$. The relative changes in mass and radius for the IOPB-I are shown in Fig. \ref{fig:bubble}. It is observed that the quarkyonic stars with (0.3, 1400) and (0.3, 800) give the largest increment in mass as well as radius i.e.(18.13\%, 21.34\%) and (16.27\%, 21.2\%). An increment in transition density lowers the mass and radius which is because the appearance of quarks at higher densities lowers the magnitude of the speed of sound. The 0.03\% dark matter case is particularly interesting as it gives an intermediate-mass range and increment of (10.2\%, 12.9\%) in mass and radius. Similar behavior is observed in the case of the G3 parameter set. We have (20.27\%, 23.2\%) of mass and radius increment for (0.3, 800), while (0.3, 800, 0.03) combination gives an increment of (12.3\%, 13.9\%) for mass and radius.

The addition of dark matter (DM) to certain models enhances their predictive capabilities, particularly in estimating the maximum mass and canonical radius of pulsars. This adjustment aligns well with observational data, such as that from notable events like GW190814 and NICER observation. The observation of GW190814 indicates a compact binary coalescence involving a black hole with a mass between 22.2 and 24.3 $M_\odot$  and a compact object weighing between 2.50 and 2.67 $M_\odot$ where all measurements quoted at the $90\%$  credible level \cite{Abbott_2020}. Its secondary component is speculated to be either the lightest black hole or the heaviest neutron star. The mass constraint of black widow pulsar PSR J0952-0607 and J0740+6620  are  M = 2.35 $\pm$ 0.17 $M_\odot$ and 2.08 $\pm$ 0.07 $M_\odot$ respectively \cite{Romani_2022, Cromartie_2020}. The radius constraint of the latter pulsar is 12.35 $\pm$ 0.35 km (revised NICER) while for a canonical star, it is 11.80 $<$ $R_{1.4}$ $<$ 13.10 km (revised NICER) \cite{Miller_2021}.  Two NICER results of  PSR J0030+0451 imply that the radius of the canonical stars must be in the range 11.52 $<$ $R_{1.4}$ $<$ 13.85 km and, and 11.96$<$ $R_{1.4}$ $<$ 14.26 km (NICER) \cite{Riley_2019, Miller_2019}. Some model variations, such as IOPB-I with higher dark matter percentages, may yield predictions that do not match observational data. Nevertheless, by exploring different combinations of free parameters within the models and employing Bayesian analysis, it's possible to refine these predictions and estimate the values of all free parameters using various observables, which may be one of our future works \cite{futurework}. 

\begin{figure*}
\centering
\includegraphics[width=0.5\columnwidth]{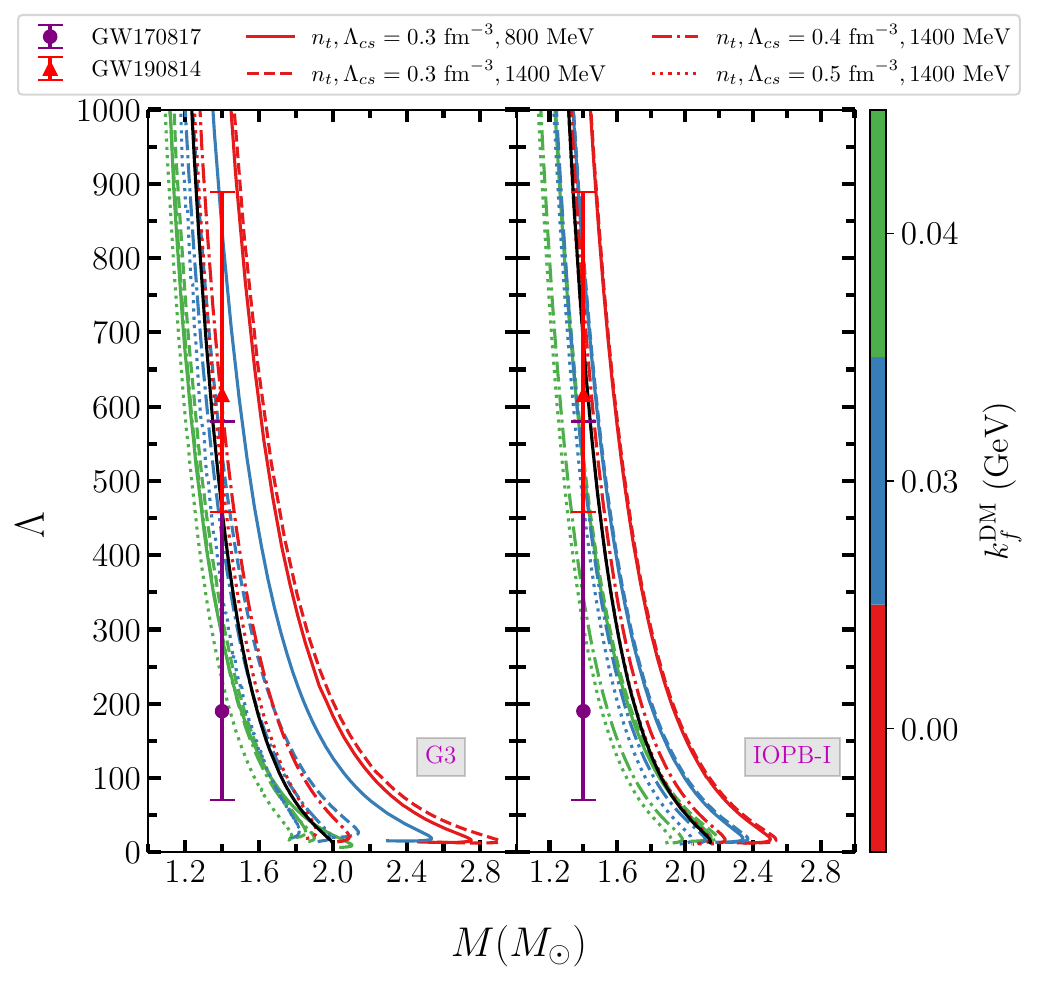}
\includegraphics[width=0.5\columnwidth]{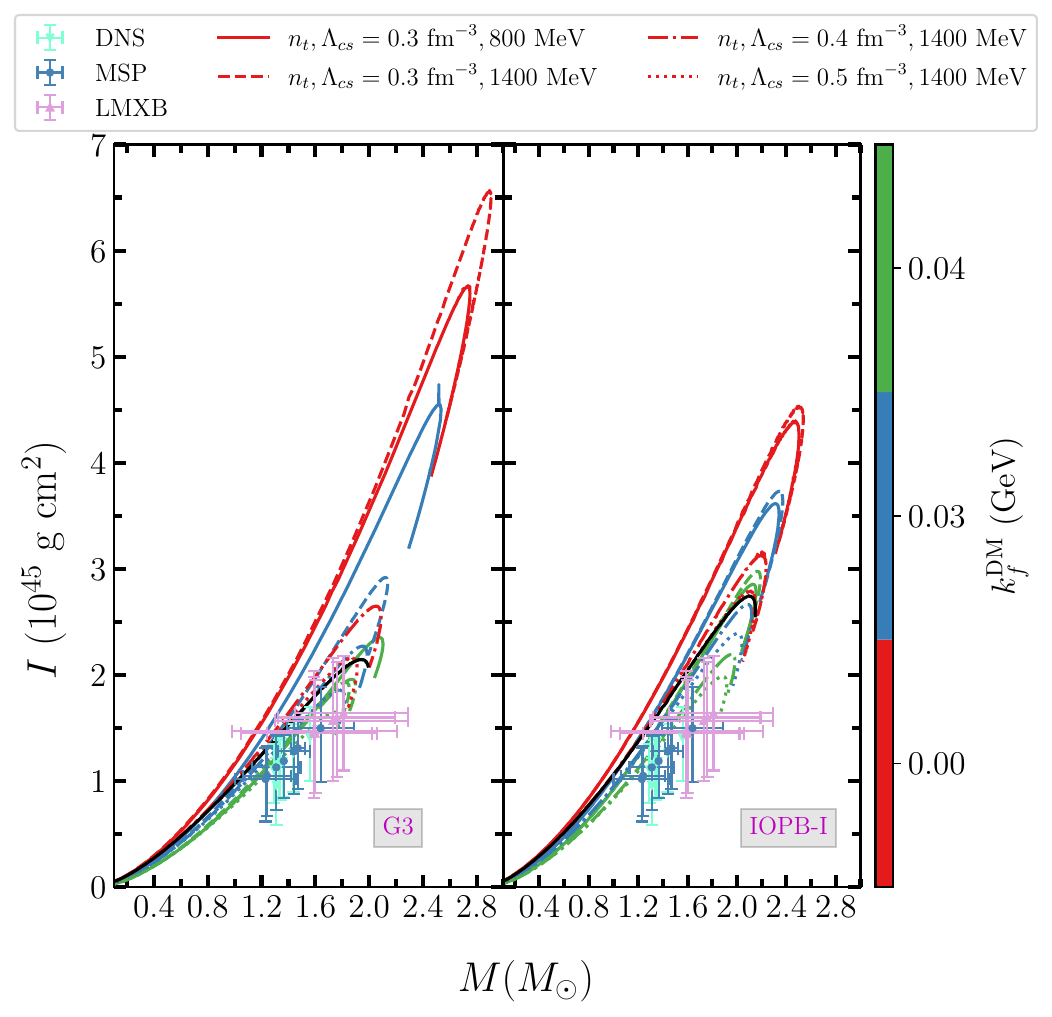}
\caption{{\it Left panel:}} The dimensionless tidal deformability ($\Lambda$) as a function of mass for different EOSs. The purple and cyan colored error bars are the observational constraints given by LIGO/Virgo (events GW170817, NS-NS merger \cite{PhysRevLett.119.161101}, and GW190814, BH-NS merger, \cite{Abbott_2020}). {\it  Right panel:} The moment of inertia of the NS as a function of mass \cite{Landry_2019}.
\label{fig:Lambda_mom}
\end{figure*}
\subsection{Tidal Deformability, and Moment of Inertia}

Tidal deformability is a key parameter that provides valuable insights into the internal structure of compact celestial objects. It enables us to understand how these objects respond to tidal forces and is essential for constraining the density at which the transition of QM takes place in the NSs. In our investigation, we explore how tidal deformability is influenced by the maximum mass of the star, including quarkyonic and DM admixed quarkyonic EOSs. Tidal deformability $\Lambda$ which quantifies the NS's ability to deform in response to an external gravitational tidal field is mathematically defined as $\Lambda = \frac{\lambda}{M^5}$. Here, $\lambda$ is the tidal deformability and can be expressed in terms of the dimensionless quadrupole tidal Love number, $k_2$, and the star's radius, given by $\lambda = \frac{2}{3} k_{2} R^{5}$ \cite{Hinderer_2008,Hinderer_2009} It's worth noting that the tidal Love number, $k_2$, is intimately connected to the internal structure and composition of the star and possibly can be measured in gravitational waves emanating from inspiralling binary NSs \cite{ref1_k2, ref2_k2}.  However, it's important to recognize that the presence of DM inside the quarkyonic star also plays a role in regulating dimensionless tidal deformability. This effect is observed as a decrease in $\Lambda$ with an increase in the mass of the star, which can be seen from the left panel of Fig. \ref{fig:Lambda_mom}. Additionally, we represent the lower and upper bounds for the tidal deformability of the secondary component in the GW170817 and GW190814 merger events for the canonical stars as a representative case for softer and stiffer EOSs. These bounds indicate the extreme values of $\Lambda$ reported in the reference \cite{Abbott_2020}, specifically $\Lambda = 190^{+390}_{-120}$ and $\Lambda = 616^{+273}_{-158}$, which provide constraints on the tidal deformability within the context of these gravitational wave events. Although it is worth emphasizing that the nature of the mass-gap object in the GW190814 is not known yet, therefore such constraint is still speculative. The magnitude of $\Lambda$ directly depends on the EOSs. Softer EOS predicts a lower magnitude than the stiffer. Since the QM stiffened the EOS, the magnitude of $\Lambda$ also increases. The values for $\Lambda_{1.4}$ i.e. the deformability at the canonical mass, are given in Table \ref{tab:all parameter set}. Only IOPB-I model parameter for $n_t, \Lambda = 0.3$ fm$^{-3}$, $1400$ MeV without dark matter case satisfies the GW190814 limit. Almost all cases satisfy the GW170817 limit for both G3 and IOPB-I models for various dark matter percentages. Therefore, the DM plays a significant role in such cases and our model is better suited for predicting properties of NSs excluding mass-gap objects. One can also put direct constraints on the amount of DM inside the NS.

The moment of inertia (MOI) of the NS depends on its mass distribution within the star and its microscopic degrees of freedom. The NSs often have a core composed of extremely dense neutron matter surrounded by a thin crust. As of now, we haven't observed the MOI for any NS. However, one can estimate the MOI of the NS using different universal relations. One such study can be found in Refs. \cite{Landry_2018, Landry_2019} for various systems that involve Millisecond pulses (MSP), Double Neutron Star (DNS) systems, and Low Mass X-ray Binaries (LMXB). Therefore, it is a crucial quantity to constrain the internal structure of the NS. In this calculation, we obtain the MOI for the slow-rotating case of DM-admixed quarkyonic star. The relation between $I-M$ is shown in the right panel of Fig. \ref{fig:Lambda_mom}. Also, the canonical MOI is enumerated in Table \ref{tab:all parameter set}. It is noticed that the DM composition significantly affects its value by softening the quarkyonic EOS. Thus, the presence of DM decelerates the rotation of the NS. Also, due to the stiffness of G3 EOS, for a fixed value of DM, the MOI is larger for G3 as compared to IOPB-I. Almost all configurations satisfy the predicted MOI for the different pulsars, mainly in that mass range. Therefore, we can put direct constraints on the free parameters used in our model with future observational data.

\section{Summary} 
\label{Conclusions}
In summary, we explore the properties of the quarkyonic stars by incorporating the influence of DM within the framework of the E-RMF formalism. This approach allows us to understand how the presence of DM affects the macroscopic properties of NSs, shedding light on the intriguing interplay between fundamental particles, dense matter, and astrophysical properties of NS. The model consists of four free parameters, such as EOS model, transition density, QCD confinement scale, and DM momentum. Neutralino is considered to be a DM candidate that interacts with the nucleons via exchanging standard model Higgs through Yukawa potential. 

The results of the theoretical simulation suggest that even though quarkyonic EOS predicts massive neutron stars greater than $2M_\odot$, the presence of DM can significantly soften the EOS, and hence, the DM-admixed NS will fall into compact stars with the smallest mass range. The primary reason for this behavior can be attributed to the fact that the inclusion of DM  increases the energy content of the quarkyonic star without affecting the pressure content of the system. Our study revealed that the maximum mass of this DM-admixed quarkyonic star should be in the range $2M_\odot < M <2.5M_\odot$. The results disclose some interesting correlations between DM parameters and various NS properties. 

A more in-depth investigation is necessary to understand the ramifications of DM, particularly its effects on the cooling process and nuclear symmetry energy in baryonic matter. Identifying stars with similar masses but varying surface temperatures may suggest that the cooler ones could be the stars infused with DM. The incorporation of DM resulted in a reduction in the radius and tidal deformability across all mass ranges, indicating that DM has a discernible impact on the structural properties of NSs. 
One can fix the range of the parameter space by incorporating the Bayesian analysis with different observational data available to date. Also, the universal relations can be useful to constrain other quantities, such as the MOI and the oscillation frequency of the quarkyonic star admixed with the DM. This might give enough information regarding the gravitational waves emitted during the inspiral and merger stages, which could be detected with terrestrial detectors and the Einstein telescope in the near future.    
\section{Acknowledgments}
JAP and AK acknowledge IOP and DST (grant no. CRG/2019/002691) and project CAS E3545KU2 for providing financial support. Also, we thank Tianqi Zhao for the valuable discussions throughout the project.
\bibliography{JCAP}

\providecommand{\href}[2]{#2}\begingroup\raggedright\begin{thebibliography}{100}

\bibitem{doi:10.1126/science.1090720}
J.M.~Lattimer and M.~Prakash, \emph{The physics of neutron stars},
  \href{https://doi.org/10.1126/science.1090720}{\emph{Science} {\bfseries 304}
  (2004) 536}.

\bibitem{Burrows2000}
A.~Burrows, \emph{Supernova explosions in the universe},
  \href{https://doi.org/10.1038/35001501}{\emph{Nature} {\bfseries 403} (2000)
  727}.

\bibitem{doi:10.1146/annurev-nucl-102711-095018}
J.M.~Lattimer, \emph{The nuclear equation of state and neutron star masses},
  \href{https://doi.org/10.1146/annurev-nucl-102711-095018}{\emph{Annual Review
  of Nuclear and Particle Science} {\bfseries 62} (2012) 485}.

\bibitem{2016ARA&A..54..401O}
F.~\"Ozel and P.~Freire, \emph{{Masses, Radii, and the Equation of State of
  Neutron Stars}},
  \href{https://doi.org/10.1146/annurev-astro-081915-023322}{\emph{araa}
  {\bfseries 54} (2016) 401}
  [\href{https://arxiv.org/abs/1603.02698}{{\ttfamily 1603.02698}}].

\bibitem{PhysRevLett.119.161101}
{\scshape LIGO Scientific Collaboration and Virgo Collaboration} collaboration,
  \emph{Gw170817: Observation of gravitational waves from a binary neutron star
  inspiral}, \href{https://doi.org/10.1103/PhysRevLett.119.161101}{\emph{Phys.
  Rev. Lett.} {\bfseries 119} (2017) 161101}.

\bibitem{PhysRevLett.121.091102}
S.~De, D.~Finstad, J.M.~Lattimer, D.A.~Brown, E.~Berger and C.M.~Biwer,
  \emph{Tidal deformabilities and radii of neutron stars from the observation
  of gw170817},
  \href{https://doi.org/10.1103/PhysRevLett.121.091102}{\emph{Phys. Rev. Lett.}
  {\bfseries 121} (2018) 091102}.

\bibitem{PhysRevLett.121.161101}
{\scshape The LIGO Scientific Collaboration and the Virgo Collaboration}
  collaboration, \emph{Gw170817: Measurements of neutron star radii and
  equation of state},
  \href{https://doi.org/10.1103/PhysRevLett.121.161101}{\emph{Phys. Rev. Lett.}
  {\bfseries 121} (2018) 161101}.

\bibitem{Capano2020}
C.D.~Capano, I.~Tews, S.M.~Brown, B.~Margalit, S.~De, S.~Kumar et~al.,
  \emph{Stringent constraints on neutron-star radii from multimessenger
  observations and nuclear theory},
  \href{https://doi.org/10.1038/s41550-020-1014-6}{\emph{Nature Astronomy}
  {\bfseries 4} (2020) 625}.

\bibitem{Riley_2019}
T.E.~{Riley}, A.L.~{Watts}, S.~{Bogdanov} et~al., \emph{{A NICER View of PSR
  J0030+0451: Millisecond Pulsar Parameter Estimation}},
  \href{https://doi.org/10.3847/2041-8213/ab481c}{\emph{APJL} {\bfseries 887}
  (2019) L21}.

\bibitem{Miller_2019}
M.C.~Miller, F.K.~Lamb, A.J.~Dittmann and other, \emph{Psr j0030+0451 mass and
  radius from nicer data and implications for the properties of neutron star
  matter}, \href{https://doi.org/10.3847/2041-8213/ab50c5}{\emph{The
  Astrophysical Journal Letters} {\bfseries 887} (2019) L24}.

\bibitem{PhysRevLett.122.122701}
L.~McLerran and S.~Reddy, \emph{Quarkyonic matter and neutron stars},
  \href{https://doi.org/10.1103/PhysRevLett.122.122701}{\emph{Phys. Rev. Lett.}
  {\bfseries 122} (2019) 122701}.

\bibitem{BURGIO200219}
G.~Burgio, M.~Baldo, P.~Sahu, A.~Santra and H.-J.~Schulze, \emph{Maximum mass
  of neutron stars with a quark core},
  \href{https://doi.org/https://doi.org/10.1016/S0370-2693(01)01479-4}{\emph{Physics
  Letters B} {\bfseries 526} (2002) 19}.

\bibitem{PhysRevC.60.025801}
K.~Schertler, S.~Leupold and J.~Schaffner-Bielich, \emph{Neutron stars and
  quark phases in the nambu--jona-lasinio model},
  \href{https://doi.org/10.1103/PhysRevC.60.025801}{\emph{Phys. Rev. C}
  {\bfseries 60} (1999) 025801}.

\bibitem{PhysRevD.102.023021}
T.~Zhao and J.M.~Lattimer, \emph{Quarkyonic matter equation of state in
  beta-equilibrium},
  \href{https://doi.org/10.1103/PhysRevD.102.023021}{\emph{Phys. Rev. D}
  {\bfseries 102} (2020) 023021}.

\bibitem{DM5}
J.M.~Cline, P.~Scott, K.~Kainulainen and C.~Weniger, \emph{Update on scalar
  singlet dark matter},
  \href{https://doi.org/10.1103/PhysRevD.88.055025}{\emph{Phys. Rev. D}
  {\bfseries 88} (2013) 055025}.

\bibitem{DM6}
G.~Bertone and M.~Fairbairn, \emph{Compact stars as dark matter probes},
  \href{https://doi.org/10.1103/PhysRevD.77.043515}{\emph{Phys. Rev. D}
  {\bfseries 77} (2008) 043515}.

\bibitem{DM7}
N.F.~Bell, G.~Busoni, T.F.~Motta, S.~Robles, A.W.~Thomas and M.~Virgato,
  \emph{Nucleon structure and strong interactions in dark matter capture in
  neutron stars},
  \href{https://doi.org/10.1103/PhysRevLett.127.111803}{\emph{Phys. Rev. Lett.}
  {\bfseries 127} (2021) 111803}.

\bibitem{DM8}
B.~Kain, \emph{Dark matter admixed neutron stars},
  \href{https://doi.org/10.1103/PhysRevD.103.043009}{\emph{Phys. Rev. D}
  {\bfseries 103} (2021) 043009}.

\bibitem{DM9}
D.~Rafiei~Karkevandi, S.~Shakeri, V.~Sagun and O.~Ivanytskyi, \emph{Bosonic
  dark matter in neutron stars and its effect on gravitational wave signal},
  \href{https://doi.org/10.1103/PhysRevD.105.023001}{\emph{Phys. Rev. D}
  {\bfseries 105} (2022) 023001}.

\bibitem{DM10}
S.P.~MARTIN, \emph{A supersymmetry primer},  vol.~Volume 18 of \emph{Advanced
  Series on Directions in High Energy Physics}, pp.~1--98, WORLD SCIENTIFIC
  (1998), \href{https://doi.org/10.1142/9789812839657_0001}{DOI}.

\bibitem{DM11}
H.C.~Das, A.~Kumar and S.K.~Patra, \emph{{Effects of dark matter on the
  in-spiral properties of the binary neutron stars}},
  \href{https://doi.org/10.1093/mnras/stab2387}{\emph{Monthly Notices of the
  Royal Astronomical Society} {\bfseries 507} (2021) 4053}.

\bibitem{DM12}
H.C.~Das, A.~Kumar, B.~Kumar and S.K.~Patra, \emph{Dark matter effects on the
  compact star properties},
  \href{https://doi.org/10.3390/galaxies10010014}{\emph{Galaxies} {\bfseries
  10} (2022) }.

\bibitem{DM13}
H.C.~Das, A.~Kumar, S.K.~Biswal and S.K.~Patra, \emph{Impacts of dark matter on
  the $f$-mode oscillation of hyperon star},
  \href{https://doi.org/10.1103/PhysRevD.104.123006}{\emph{Phys. Rev. D}
  {\bfseries 104} (2021) 123006}.

\bibitem{routray}
P.~Routaray, H.C.~Das, S.~Sen, B.~Kumar, G.~Panotopoulos and T.~Zhao,
  \emph{Radial oscillations of dark matter admixed neutron stars},
  \href{https://doi.org/10.1103/PhysRevD.107.103039}{\emph{Phys. Rev. D}
  {\bfseries 107} (2023) 103039}.

\bibitem{PhysRevD.83.083512}
C.~Kouvaris and P.~Tinyakov, \emph{Constraining asymmetric dark matter through
  observations of compact stars},
  \href{https://doi.org/10.1103/PhysRevD.83.083512}{\emph{Phys. Rev. D}
  {\bfseries 83} (2011) 083512}.

\bibitem{DM2}
A.~Quddus, G.~Panotopoulos, B.~Kumar, S.~Ahmad and S.K.~Patra, \emph{Gw170817
  constraints on the properties of a neutron star in the presence of wimp dark
  matter}, \href{https://doi.org/10.1088/1361-6471/ab9d36}{\emph{Journal of
  Physics G: Nuclear and Particle Physics} {\bfseries 47} (2020) 095202}.

\bibitem{2017IJMPA..3230023B}
N.~{Bernal}, M.~{Heikinheimo}, T.~{Tenkanen}, K.~{Tuominen} and V.~{Vaskonen},
  \emph{{The dawn of FIMP Dark Matter: A review of models and constraints}},
  \href{https://doi.org/10.1142/S0217751X1730023X}{\emph{International Journal
  of Modern Physics A} {\bfseries 32} (2017) 1730023}.

\bibitem{Hall2010}
L.J.~Hall, K.~Jedamzik, J.~March-Russell and S.M.~West, \emph{Freeze-in
  production of fimp dark matter},
  \href{https://doi.org/10.1007/JHEP03(2010)080}{\emph{Journal of High Energy
  Physics} {\bfseries 2010} (2010) 80}.

\bibitem{PhysRevD.69.035001}
D.~Hooper and L.-T.~Wang, \emph{Direct and indirect detection of neutralino
  dark matter in selected supersymmetry breaking scenarios},
  \href{https://doi.org/10.1103/PhysRevD.69.035001}{\emph{Phys. Rev. D}
  {\bfseries 69} (2004) 035001}.

\bibitem{2014JHEP...08..093H}
T.~{Han}, Z.~{Liu} and S.~{Su}, \emph{{Light neutralino dark matter:
  direct/indirect detection and collider searches}},
  \href{https://doi.org/10.1007/JHEP08(2014)093}{\emph{Journal of High Energy
  Physics} {\bfseries 2014} (2014) 93}.

\bibitem{PhysRevD.99.043016}
A.~Das, T.~Malik and A.C.~Nayak, \emph{Confronting nuclear equation of state in
  the presence of dark matter using gw170817 observation in relativistic mean
  field theory approach},
  \href{https://doi.org/10.1103/PhysRevD.99.043016}{\emph{Phys. Rev. D}
  {\bfseries 99} (2019) 043016}.

\bibitem{Duffy_2009}
L.D.~Duffy and K.~van Bibber, \emph{Axions as dark matter particles},
  \href{https://doi.org/10.1088/1367-2630/11/10/105008}{\emph{New Journal of
  Physics} {\bfseries 11} (2009) 105008}.

\bibitem{dark_matter_3}
G.~Busoni, \emph{Capture of dark matter in neutron stars},
  \href{https://doi.org/10.3103/S0027134922020205}{\emph{Moscow University
  Physics Bulletin} {\bfseries 77} (2022) 301}.

\bibitem{dark_matter_4}
N.~Raj, P.~Tanedo and H.-B.~Yu, \emph{Neutron stars at the dark matter direct
  detection frontier},
  \href{https://doi.org/10.1103/PhysRevD.97.043006}{\emph{Phys. Rev. D}
  {\bfseries 97} (2018) 043006}.

\bibitem{Bernabei2008}
R.~Bernabei, P.~Belli, F.~Cappella et~al., \emph{First results from dama/libra
  and the combined results with dama/nai},
  \href{https://doi.org/10.1140/epjc/s10052-008-0662-y}{\emph{The European
  Physical Journal C} {\bfseries 56} (2008) 333}.

\bibitem{Bernabei2010}
R.~Bernabei, P.~Belli, F.~Cappella et~al., \emph{New results from dama/libra},
  \href{https://doi.org/10.1140/epjc/s10052-010-1303-9}{\emph{The European
  Physical Journal C} {\bfseries 67} (2010) 39}.

\bibitem{PhysRevLett.101.091301}
{\scshape XENON10 Collaboration} collaboration, \emph{Limits on spin-dependent
  wimp-nucleon cross sections from the xenon10 experiment},
  \href{https://doi.org/10.1103/PhysRevLett.101.091301}{\emph{Phys. Rev. Lett.}
  {\bfseries 101} (2008) 091301}.

\bibitem{doi:10.1126/science.1186112}
T.C.I.~Collaboration, \emph{Dark matter search results from the cdms ii
  experiment}, \href{https://doi.org/10.1126/science.1186112}{\emph{Science}
  {\bfseries 327} (2010) 1619}.

\bibitem{conrad2014indirect}
J.~Conrad, \emph{Indirect detection of wimp dark matter: a compact review},
  2014.

\bibitem{doi:10.1080/14786435608238186}
T.H.R.~Skyrme, \emph{Cvii. the nuclear surface},
  \href{https://doi.org/10.1080/14786435608238186}{\emph{The Philosophical
  Magazine: A Journal of Theoretical Experimental and Applied Physics}
  {\bfseries 1} (1956) 1043}.

\bibitem{SKYRME1958615}
T.~Skyrme, \emph{The effective nuclear potential},
  \href{https://doi.org/https://doi.org/10.1016/0029-5582(58)90345-6}{\emph{Nuclear
  Physics} {\bfseries 9} (1958) 615}.

\bibitem{PhysRevC.5.626}
D.~Vautherin and D.M.~Brink, \emph{Hartree-fock calculations with skyrme's
  interaction. i. spherical nuclei},
  \href{https://doi.org/10.1103/PhysRevC.5.626}{\emph{Phys. Rev. C} {\bfseries
  5} (1972) 626}.

\bibitem{CHABANAT1998231}
E.~Chabanat, P.~Bonche, P.~Haensel, J.~Meyer and R.~Schaeffer, \emph{A skyrme
  parametrization from subnuclear to neutron star densities part ii. nuclei far
  from stabilities},
  \href{https://doi.org/https://doi.org/10.1016/S0375-9474(98)00180-8}{\emph{Nuclear
  Physics A} {\bfseries 635} (1998) 231}.

\bibitem{PhysRevC.58.220}
B.~Alex~Brown, \emph{New skyrme interaction for normal and exotic nuclei},
  \href{https://doi.org/10.1103/PhysRevC.58.220}{\emph{Phys. Rev. C} {\bfseries
  58} (1998) 220}.

\bibitem{STONE2007587}
J.~Stone and P.-G.~Reinhard, \emph{The skyrme interaction in finite nuclei and
  nuclear matter},
  \href{https://doi.org/https://doi.org/10.1016/j.ppnp.2006.07.001}{\emph{Progress
  in Particle and Nuclear Physics} {\bfseries 58} (2007) 587}.

\bibitem{PhysRevC.85.035201}
M.~Dutra, O.~Louren\ifmmode~\mbox{\c{c}}\else \c{c}\fi{}o, J.S.~S\'a~Martins,
  A.~Delfino, J.R.~Stone and P.D.~Stevenson, \emph{Skyrme interaction and
  nuclear matter constraints},
  \href{https://doi.org/10.1103/PhysRevC.85.035201}{\emph{Phys. Rev. C}
  {\bfseries 85} (2012) 035201}.

\bibitem{PhysRevC.21.1568}
J.~Decharg\'e and D.~Gogny, \emph{Hartree-fock-bogolyubov calculations with the
  $d1$ effective interaction on spherical nuclei},
  \href{https://doi.org/10.1103/PhysRevC.21.1568}{\emph{Phys. Rev. C}
  {\bfseries 21} (1980) 1568}.

\bibitem{PhysRevC.63.044303}
M.~Rashdan, \emph{Structure of exotic nuclei and superheavy elements in a
  relativistic shell model},
  \href{https://doi.org/10.1103/PhysRevC.63.044303}{\emph{Phys. Rev. C}
  {\bfseries 63} (2001) 044303}.

\bibitem{PhysRevC.97.045806}
B.~Kumar, S.K.~Patra and B.K.~Agrawal, \emph{New relativistic effective
  interaction for finite nuclei, infinite nuclear matter, and neutron stars},
  \href{https://doi.org/10.1103/PhysRevC.97.045806}{\emph{Phys. Rev. C}
  {\bfseries 97} (2018) 045806}.

\bibitem{patt1}
J.A.~Pattnaik, M.~Bhuyan, R.N.~Panda and S.K.~Patra, \emph{Isotopic shift in
  magic nuclei within relativistic mean-field formalism},
  \href{https://doi.org/10.1088/1402-4896/ac3a4d}{\emph{Physica Scripta}
  {\bfseries 96} (2021) 125319}.

\bibitem{patt2}
J.A.~Pattnaik, J.T.~Majekodunmi, A.~Kumar, M.~Bhuyan and S.K.~Patra,
  \emph{Appearance of a peak in the symmetry energy at $n=126$ for the pb
  isotopic chain within the relativistic energy density functional approach},
  \href{https://doi.org/10.1103/PhysRevC.105.014318}{\emph{Phys. Rev. C}
  {\bfseries 105} (2022) 014318}.

\bibitem{patt3}
J.A.~Pattnaik, R.N.~Panda, M.~Bhuyan and S.K.~Patra, \emph{Constraining the
  relativistic mean-field models from prex-2 data: effective forces revisited
  *}, \href{https://doi.org/10.1088/1674-1137/ac6f4e}{\emph{Chinese Physics C}
  {\bfseries 46} (2022) 094103}.

\bibitem{patt4}
J.A.~Pattnaik, K.C.~Naik, R.N.~Panda, M.~Bhuyan and S.K.~Patra, \emph{Structure
  and reaction studies of {\$}{\$}z=120{\$}{\$}isotopes using non-relativistic
  and relativistic mean-field formalisms},
  \href{https://doi.org/10.1007/s12043-023-02619-9}{\emph{Pramana} {\bfseries
  97} (2023) 136}.

\bibitem{WALECKA1974491}
J.~Walecka, \emph{A theory of highly condensed matter},
  \href{https://doi.org/https://doi.org/10.1016/0003-4916(74)90208-5}{\emph{Annals
  of Physics} {\bfseries 83} (1974) 491}.

\bibitem{BOGUTA1977413}
J.~Boguta and A.~Bodmer, \emph{Relativistic calculation of nuclear matter and
  the nuclear surface},
  \href{https://doi.org/https://doi.org/10.1016/0375-9474(77)90626-1}{\emph{Nuclear
  Physics A} {\bfseries 292} (1977) 413}.

\bibitem{Serot1992}
B.D.~Serot and J.D.~Walecka, \emph{Relativistic nuclear many-body theory},  in
  \emph{Recent Progress in Many-Body Theories: Volume 3}, T.L.~Ainsworth,
  C.E.~Campbell, B.E.~Clements and E.~Krotscheck, eds., (Boston, MA),
  pp.~49--92, Springer US (1992),
  \href{https://doi.org/10.1007/978-1-4615-3466-2_5}{DOI}.

\bibitem{SEROT1979146}
B.D.~Serot, \emph{A relativistic nuclear field theory with $\pi$ and $\rho$
  mesons},
  \href{https://doi.org/https://doi.org/10.1016/0370-2693(79)90804-9}{\emph{Physics
  Letters B} {\bfseries 86} (1979) 146}.

\bibitem{DM1}
G.~Panotopoulos and I.~Lopes, \emph{Dark matter effect on realistic equation of
  state in neutron stars},
  \href{https://doi.org/10.1103/PhysRevD.96.083004}{\emph{Phys. Rev. D}
  {\bfseries 96} (2017) 083004}.

\bibitem{PhysRevC.55.540}
G.A.~Lalazissis, J.~K\"onig and P.~Ring, \emph{New parametrization for the
  lagrangian density of relativistic mean field theory},
  \href{https://doi.org/10.1103/PhysRevC.55.540}{\emph{Phys. Rev. C} {\bfseries
  55} (1997) 540}.

\bibitem{PhysRevC.74.045806}
A.~Sulaksono and T.~Mart, \emph{Low density instability in relativistic mean
  field models}, \href{https://doi.org/10.1103/PhysRevC.74.045806}{\emph{Phys.
  Rev. C} {\bfseries 74} (2006) 045806}.

\bibitem{PhysRevC.70.058801}
D.P.~Menezes and C.~Provid\^encia, \emph{$\ensuremath{\delta}$ meson effects on
  stellar matter},
  \href{https://doi.org/10.1103/PhysRevC.70.058801}{\emph{Phys. Rev. C}
  {\bfseries 70} (2004) 058801}.

\bibitem{LALAZISSIS200936}
G.~Lalazissis, S.~Karatzikos, R.~Fossion, D.P.~Arteaga, A.~Afanasjev and
  P.~Ring, \emph{The effective force nl3 revisited},
  \href{https://doi.org/https://doi.org/10.1016/j.physletb.2008.11.070}{\emph{Physics
  Letters B} {\bfseries 671} (2009) 36}.

\bibitem{PhysRevC.82.055803}
F.J.~Fattoyev, C.J.~Horowitz, J.~Piekarewicz and G.~Shen, \emph{Relativistic
  effective interaction for nuclei, giant resonances, and neutron stars},
  \href{https://doi.org/10.1103/PhysRevC.82.055803}{\emph{Phys. Rev. C}
  {\bfseries 82} (2010) 055803}.

\bibitem{PhysRevC.82.025203}
A.b.A.~Dadi, \emph{Parametrization of the relativistic
  $\ensuremath{\sigma}$-$\ensuremath{\omega}$ model for nuclear matter},
  \href{https://doi.org/10.1103/PhysRevC.82.025203}{\emph{Phys. Rev. C}
  {\bfseries 82} (2010) 025203}.

\bibitem{PhysRevC.84.054309}
X.~Roca-Maza, X.~Vi\~nas, M.~Centelles, P.~Ring and P.~Schuck,
  \emph{Relativistic mean-field interaction with density-dependent
  meson-nucleon vertices based on microscopical calculations},
  \href{https://doi.org/10.1103/PhysRevC.84.054309}{\emph{Phys. Rev. C}
  {\bfseries 84} (2011) 054309}.

\bibitem{PhysRevC.85.024302}
B.-J.~Cai and L.-W.~Chen, \emph{Nuclear matter fourth-order symmetry energy in
  the relativistic mean field models},
  \href{https://doi.org/10.1103/PhysRevC.85.024302}{\emph{Phys. Rev. C}
  {\bfseries 85} (2012) 024302}.

\bibitem{PhysRevC.102.065805}
F.J.~Fattoyev, C.J.~Horowitz, J.~Piekarewicz and B.~Reed, \emph{Gw190814:
  Impact of a 2.6 solar mass neutron star on the nucleonic equations of state},
  \href{https://doi.org/10.1103/PhysRevC.102.065805}{\emph{Phys. Rev. C}
  {\bfseries 102} (2020) 065805}.

\bibitem{E-RMF1}
H.~Das, A.~Kumar, B.~Kumar, S.~Biswal and S.~Patra, \emph{Impacts of dark
  matter on the curvature of the neutron star},
  \href{https://doi.org/10.1088/1475-7516/2021/01/007}{\emph{Journal of
  Cosmology and Astroparticle Physics} {\bfseries 2021} (2021) 007}.

\bibitem{E-RMF2}
S.K.~Patra, M.~Centelles, X.~Vi\~nas and M.~Del~Estal, \emph{Surface
  incompressibility from semiclassical relativistic mean field calculations},
  \href{https://doi.org/10.1103/PhysRevC.65.044304}{\emph{Phys. Rev. C}
  {\bfseries 65} (2002) 044304}.

\bibitem{E-RMF3}
H.~Müller and B.D.~Serot, \emph{Relativistic mean-field theory and the
  high-density nuclear equation of state},
  \href{https://doi.org/https://doi.org/10.1016/0375-9474(96)00187-X}{\emph{Nuclear
  Physics A} {\bfseries 606} (1996) 508}.

\bibitem{E-RMF4}
P.~Wang, \emph{Asymmetric nuclear matter at finite temperature and density},
  \href{https://doi.org/10.1103/PhysRevC.61.054904}{\emph{Phys. Rev. C}
  {\bfseries 61} (2000) 054904}.

\bibitem{E-RMF5}
A.~Kumar, H.C.~Das, S.K.~Biswal, B.~Kumar and S.K.~Patra, \emph{Warm dense
  matter and cooling of supernovae remnants},
  \href{https://doi.org/10.1140/epjc/s10052-020-8353-4}{\emph{The European
  Physical Journal C} {\bfseries 80} (2020) 775}.

\bibitem{Reinhard_1989}
P.G.~Reinhard, \emph{The relativistic mean-field description of nuclei and
  nuclear dynamics},
  \href{https://doi.org/10.1088/0034-4885/52/4/002}{\emph{Reports on Progress
  in Physics} {\bfseries 52} (1989) 439}.

\bibitem{KUMAR2017197}
B.~Kumar, S.~Singh, B.~Agrawal and S.~Patra, \emph{New parameterization of the
  effective field theory motivated relativistic mean field model},
  \href{https://doi.org/https://doi.org/10.1016/j.nuclphysa.2017.07.001}{\emph{Nuclear
  Physics A} {\bfseries 966} (2017) 197}.

\bibitem{ERMF6}
L.D.~Miller and A.E.S.~Green, \emph{Relativistic self-consistent meson field
  theory of spherical nuclei},
  \href{https://doi.org/10.1103/PhysRevC.5.241}{\emph{Phys. Rev. C} {\bfseries
  5} (1972) 241}.

\bibitem{ERMF7}
R.J.~Furnstahl, C.E.~Price and G.E.~Walker, \emph{Systematics of light deformed
  nuclei in relativistic mean-field models},
  \href{https://doi.org/10.1103/PhysRevC.36.2590}{\emph{Phys. Rev. C}
  {\bfseries 36} (1987) 2590}.

\bibitem{E-RMF8}
P.-G.~Reinhard, \emph{The nonlinearity of the scalar field in a relativistic
  mean-field theory of the nucleus},
  \href{https://doi.org/10.1007/BF01290231}{\emph{Zeitschrift f{\"u}r Physik A
  Atomic Nuclei} {\bfseries 329} (1988) 257}.

\bibitem{E-RMF9}
R.~Furnstahl, B.D.~Serot and H.-B.~Tang, \emph{A chiral effective lagrangian
  for nuclei},
  \href{https://doi.org/https://doi.org/10.1016/S0375-9474(96)00472-1}{\emph{Nuclear
  Physics A} {\bfseries 615} (1997) 441}.

\bibitem{Glendenning}
N.K.~Glendenning, \emph{{Compact stars}} (1997).

\bibitem{DM3}
A.~Das, T.~Malik and A.C.~Nayak, \emph{Confronting nuclear equation of state in
  the presence of dark matter using gw170817 observation in relativistic mean
  field theory approach},
  \href{https://doi.org/10.1103/PhysRevD.99.043016}{\emph{Phys. Rev. D}
  {\bfseries 99} (2019) 043016}.

\bibitem{mnras}
H.C.~Das, A.~Kumar, B.~Kumar, S.K.~Biswal, T.~Nakatsukasa, A.~Li et~al.,
  \emph{{Effects of dark matter on the nuclear and neutron star matter}},
  \href{https://doi.org/10.1093/mnras/staa1435}{\emph{Monthly Notices of the
  Royal Astronomical Society} {\bfseries 495} (2020) 4893}.

\bibitem{ankit_arxiv}
A.~Kumar, D.~Dey, S.~Haque, R.~Mallick and S.K.~Patra, \emph{Quarkyonic model
  for neutron star matter: A relativistic mean-field approach},  2023.

\bibitem{Perkins}
D.H.~Perkins, \emph{{Particle astrophysics}} (2003).

\bibitem{Parmar_2022}
V.~Parmar, H.C.~Das, A.~Kumar, A.~Kumar, M.K.~Sharma, P.~Arumugam et~al.,
  \emph{Pasta properties of the neutron star within effective relativistic
  mean-field model},
  \href{https://doi.org/10.1103/PhysRevD.106.023031}{\emph{Phys. Rev. D}
  {\bfseries 106} (2022) 023031}.

\bibitem{saturation_density}
H.A.~Bethe, \emph{Theory of nuclear matter},
  \href{https://doi.org/https://doi.org/10.1146/annurev.ns.21.120171.000521}{\emph{Annual
  Review of Nuclear and Particle Science} {\bfseries 21} (1971) 93}.

\bibitem{effective_mass}
T.~Marketin, D.~Vretenar and P.~Ring, \emph{Calculation of
  \ensuremath{\beta}-decay rates in a relativistic model with
  momentum-dependent self-energies},
  \href{https://doi.org/10.1103/PhysRevC.75.024304}{\emph{Phys. Rev. C}
  {\bfseries 75} (2007) 024304}.

\bibitem{K}
U.~Garg and G.~Colò, \emph{The compression-mode giant resonances and nuclear
  incompressibility},
  \href{https://doi.org/https://doi.org/10.1016/j.ppnp.2018.03.001}{\emph{Progress
  in Particle and Nuclear Physics} {\bfseries 101} (2018) 55}.

\bibitem{Ks}
J.~Zimmerman, Z.~Carson, K.~Schumacher, A.W.~Steiner and K.~Yagi,
  \emph{Measuring nuclear matter parameters with nicer and ligo/virgo},  2020.

\bibitem{sos1}
I.~Tews, J.~Margueron and S.~Reddy, \emph{Critical examination of constraints
  on the equation of state of dense matter obtained from gw170817},
  \href{https://doi.org/10.1103/PhysRevC.98.045804}{\emph{Phys. Rev. C}
  {\bfseries 98} (2018) 045804}.

\bibitem{TOV1}
R.C.~Tolman, \emph{Static solutions of einstein's field equations for spheres
  of fluid}, \href{https://doi.org/10.1103/PhysRev.55.364}{\emph{Phys. Rev.}
  {\bfseries 55} (1939) 364}.

\bibitem{TOV2}
J.R.~Oppenheimer and G.M.~Volkoff, \emph{On massive neutron cores},
  \href{https://doi.org/10.1103/PhysRev.55.374}{\emph{Phys. Rev.} {\bfseries
  55} (1939) 374}.

\bibitem{Abbott_2020}
R.~Abbott, T.D.~Abbott, S.~Abraham, F.~Acernese, K.~Ackley, C.~Adams et~al.,
  \emph{Gw190814: Gravitational waves from the coalescence of a 23 solar mass
  black hole with a 2.6 solar mass compact object},
  \href{https://doi.org/10.3847/2041-8213/ab960f}{\emph{The Astrophysical
  Journal Letters} {\bfseries 896} (2020) L44}.

\bibitem{Romani_2022}
R.W.~Romani, D.~Kandel, A.V.~Filippenko, T.G.~Brink and W.~Zheng, \emph{Psr
  j0952-0607: The fastest and heaviest known galactic neutron star},
  \href{https://doi.org/10.3847/2041-8213/ac8007}{\emph{The Astrophysical
  Journal Letters} {\bfseries 934} (2022) L17}.

\bibitem{Cromartie_2020}
H.T.~Cromartie, E.~Fonseca, S.M.~Ransom, P.B.~Demorest, Z.~Arzoumanian et~al.,
  \emph{Relativistic shapiro delay measurements of an extremely massive
  millisecond pulsar},
  \href{https://doi.org/10.1038/s41550-019-0880-2}{\emph{Nature Astronomy}
  {\bfseries 4} (2020) 72–76}.

\bibitem{Miller_2021}
M.C.~Miller, F.K.~Lamb, A.J.~Dittmann et~al., \emph{The radius of {PSR}
  j0740+6620 from {NICER} and {XMM}-newton data},
  \href{https://doi.org/10.3847/2041-8213/ac089b}{\emph{The Astrophysical
  Journal Letters} {\bfseries 918} (2021) L28}.

\bibitem{futurework}
D.~Dey, J.A.~Pattnaik and S.K.~Patra, \emph{Under preparation},  2024.

\bibitem{Landry_2019}
B.~Kumar and P.~Landry, \emph{Inferring neutron star properties from gw170817
  with universal relations},
  \href{https://doi.org/10.1103/PhysRevD.99.123026}{\emph{Phys. Rev. D}
  {\bfseries 99} (2019) 123026}.

\bibitem{Hinderer_2008}
T.~Hinderer, \emph{Tidal love numbers of neutron stars},
  \href{https://doi.org/10.1086/533487}{\emph{The Astrophysical Journal}
  {\bfseries 677} (2008) 1216}.

\bibitem{Hinderer_2009}
T.~Hinderer, \emph{Erratum: “tidal love numbers of neutron stars” (2008,
  apj, 677, 1216)},
  \href{https://doi.org/10.1088/0004-637X/697/1/964}{\emph{The Astrophysical
  Journal} {\bfseries 697} (2009) 964}.

\bibitem{ref1_k2}
T.~Hinderer, \emph{Tidal love numbers of neutron stars},
  \href{https://doi.org/10.1086/533487}{\emph{The Astrophysical Journal}
  {\bfseries 677} (2008) 1216}.

\bibitem{ref2_k2}
T.~Binnington and E.~Poisson, \emph{Relativistic theory of tidal love numbers},
  \href{https://doi.org/10.1103/PhysRevD.80.084018}{\emph{Phys. Rev. D}
  {\bfseries 80} (2009) 084018}.

\bibitem{Landry_2018}
P.~Landry and B.~Kumar, \emph{Constraints on the moment of inertia of psr
  j0737-3039a from gw170817},
  \href{https://doi.org/10.3847/2041-8213/aaee76}{\emph{The Astrophysical
  Journal Letters} {\bfseries 868} (2018) L22}.

\end{thebibliography}\endgroup
\bibliographystyle{JHEP}
\end{document}